\documentclass[journal,letterpaper]{IEEEtran}

\usepackage{cite}

\usepackage[cmex10]{amsmath}
\usepackage[latin2]{inputenc}
\usepackage{t1enc}
\usepackage{amssymb}
\usepackage{amsmath}
\usepackage[mathscr]{eucal}
\usepackage{amsthm}
\usepackage{array}
\usepackage{color}
\usepackage{amscd}

\bibliography{IEEEabrv,mybibfile}

\newtheorem{thm}{Theorem}[section]
\newtheorem{prop}[thm]{Proposition}
\newtheorem{lemma}[thm]{Lemma}
\newtheorem{cor}[thm]{Corollary}
\newtheorem{remark}[thm]{Remark}
\newtheorem{rem}[thm]{Remark}

\def\cH{\mathcal{H}}
\def\cS{\mathcal{S}}
\def\cK{\mathcal{K}}

\def\bR{\mathbb{R}}
\def\eps{\varepsilon}
\def\<{\langle}
\def\>{\rangle}

\def\vfi{\varphi}
\def\hil{{\mathcal H}}

\def\K{\mathcal{K}}

\def\B{{\mathcal B}}
\def\C{{\mathcal C}}

\def\X{{\mathcal X}}

\def\S{{\mathcal S}}
\def\M{\mathcal{M}}

\def\half{\frac{1}{2}}

\def\ep{\varepsilon}
\def\N{\mathbb{N}}
\def\iC{\mathbb{C}}
\def\R{\mathbb{R}}

\def\bz{\left(}
\def\jz{\right)}

\def\Exp{\mathbb{E}}
\def\map{\Phi}

\def\ch{W}

\newcommand{\ki}{\emph}
\newcommand{\kii}[1]{``#1''}
\newcommand{\kiii}{\emph}

\newcommand{\s}{\mbox{ }}
\newcommand{\ds}{\mbox{ }\mbox{ }}

\newcommand{\norm}[1]{\left\| #1\right\|}

\newcommand{\abs}[1]{\left| #1 \right|}

\newcommand{\diad}[2]{|#1\rangle\langle #2|}
\newcommand{\pr}[1]{\diad{#1}{#1}}
\newcommand{\D}{\hat}

\newcommand{\sr}[2]{S\bz #1\,||\, #2\jz}

\newcommand{\rsr}[3]{S_{#3}\bz #1\,||\, #2\jz}
\newcommand{\qrsr}[3]{Q_{#3}\bz #1\,||\, #2\jz}
\newcommand{\cutoff}[3]{C_{#3}\bz #1\,||\, #2\jz}
\newcommand{\srmax}[2]{S_{\mathrm{max}}\bz #1\,||\, #2\jz}

\newcommand{\hli}[3]{\underline h_{#3}\bz #1\,||\,#2\jz}
\newcommand{\hls}[3]{\overline h_{#3}\bz #1\,||\,#2\jz}
\newcommand{\hl}[3]{h_{#3}\bz #1\,||\,#2\jz}

\newcommand{\hdist}[3]{H_{#3}\bz #1\,||\,#2\jz}
\newcommand{\dist}[2]{D\bz #1\,||\,#2\jz}

\newcommand{\channel}[1]{W_{#1}}
\newcommand{\capt}[1]{\chi_{_{#1}}}

\DeclareMathOperator{\id}{id}
\DeclareMathOperator{\Tr}{Tr}
\DeclareMathOperator{\supp}{supp}

\DeclareMathOperator{\sgn}{sign}
\DeclareMathOperator{\ran}{ran}

\begin{document}
\title{On the quantum R\'enyi relative entropies and related capacity formulas}

\author{Mil\'an Mosonyi and
        Fumio Hiai
\thanks{
The work of M. Mosonyi was supported by the Hungarian Research Grant
OTKA T068258, the JSPS Japan-Hungary Joint Project, and by the 
Centre for Quantum Technologies, which is funded by the
Singapore Ministry of Education and the National Research Foundation as part of the
Research Centres of Excellence program.
Part of this work was done when M. Mosonyi was a Scientific Researcher at the Fields Institute during the Thematic Program on Mathematics in Quantum Information.
The work of F. Hiai was supported by the Grant-in-Aid for Scientific Research (C)21540208 and by the JSPS Japan-Hungary Joint Project.
}%
\thanks{M. Mosonyi is with the Centre for Quantum Technologies, National University of Singapore,
117543 Singapore and the Mathematical Institute, Budapest University of Technology and Economics
Budapest, 1111 Hungary (e-mail: milan.mosonyi@gmail.com).}
\thanks{F. Hiai is with the Graduate School of Information Sciences, Tohoku University, Sendai, 980-8579 Japan (e-mail: hiai@math.is.tohoku.ac.jp).}
}

\markboth{Submitted to the IEEE Transactions on Information Theory }%
{}

\maketitle

\begin{abstract}
Following Csisz\'ar's approach in classical information theory,
we show that the quantum $\alpha$-relative entropies with parameter $\alpha\in (0,1)$ can be represented as generalized cutoff rates, and hence provide a direct operational interpretation to the quantum $\alpha$-relative entropies. We also show that various generalizations of the Holevo capacity, defined in terms of the $\alpha$-relative entropies, coincide for the parameter range $\alpha\in (0,2]$, and show an upper bound on the one-shot $\ep$-capacity of a classical-quantum channel in terms of these capacities.
\end{abstract}

\begin{IEEEkeywords}
R\'enyi relative entropies, Hoeffding distances, generalized cutoff rates, quantum channels, $\alpha$-capacities, one-shot capacities.
\end{IEEEkeywords}

\IEEEpeerreviewmaketitle

\section{Introduction}

\IEEEPARstart{I}{n}
information theory, it is convenient to measure the distance of states (probability 
distributions in the classical, and density operators in the quantum case) with measures 
that do not satisfy the axioms of a metric. In a broad sense, a \ki{statistical distance} is 
a function taking non-negative values on pairs of states, that satisfies some convexity 
properties in its arguments and which cannot increase when its arguments are subjected to a 
stochastic operation. 
Probably the most popular statistical distance, for a good reason, is the \ki{relative 
entropy} $S$, defined for density operators $\rho,\sigma$ as
\begin{equation*}
\sr{\rho}{\sigma}:=\begin{cases}
\Tr\rho(\log\rho-\log\sigma),& \text{if } \supp\rho\le\supp\sigma,\\
 +\infty, & \text{otherwise}.
\end{cases}
\end{equation*}
While various generalizations of the relative entropy, leading to statistical distances in 
the above sense, are easy to define, they are not 
equally important, and the relevant ones are those that appear in answers to natural 
statistical problems, or in other terms, those that admit an operational interpretation.

The operational interpretation of the relative entropy is given in the problem of 
\ki{asymptotic binary state discrimination}, where one is provided with several 
identical copies of a quantum system and the knowledge that the state of the system is 
either $\rho$ (\ki{null hypothesis}) or $\sigma$ (\ki{alternative hypothesis}), where $\rho$ 
and $\sigma$ are density operators on the system's Hilbert space $\hil$, and one's goal is 
to make a good guess for the true state of the system, based on measurement results on the 
copies. It is easy to see that the most general inference scheme, based on measurements on $n$ copies, can be described by a 
binary positive operator valued measurement $(T,I-T)$, where $T\in\B(\hil^{\otimes n}),\,0\le T\le I$, 
and the guess is $\rho$ if the outcome corresponding to $T$ occurs, and $\sigma$ otherwise. 
The probability of 
a wrong guess is $\alpha_n(T):=\Tr\rho^{\otimes n}(I-T)$ if the true state is $\rho$ 
(\ki{error probability of the first kind}) and
$\beta_n(T):=\Tr\sigma^{\otimes n} T$ if the true state is $\sigma$ (\ki{error probability 
of 
the second kind}). Unless the two states have orthogonal supports, there is a trade-off 
between the two error probabilities, and it is not possible to find a measurement that makes 
both error probabilities equal to zero. 
As it turns out, if we require the error probabilities of the first kind to go to zero 
asymptotically then, under an optimal sequence of measurements, the error probabilities of 
the second kind decay exponentially, and the decay rate is given by $\sr{\rho}{\sigma}$ 
\cite{HP,ON2}. On the other hand, if we impose the stronger condition that the error probabilities of the first kind go to zero asymptotically
as $\alpha_n\sim 2^{-nr}$ for some $r>0$ then, under an optimal sequence of 
measurements, the error probabilities of 
the second kind decay as $\beta_n\sim 2^{-n\hdist{\rho}{\sigma}{r}}$, where 
$\hdist{\rho}{\sigma}{r}$ is the \ki{Hoeffding distance} of $\rho$ and $\sigma$ with 
parameter $r$ \cite{ANSzV,Hayashi,HMO2,Nagaoka}.

The Hoeffding distances can be obtained as a certain transform of the \ki{$\alpha$-relative 
entropies} that were defined by R\'enyi, based on purely axiomatic considerations 
\cite{Renyi}. While the above state discrimination result relates R\'enyi's 
$\alpha$-relative entropies to statistical distances with operational interpretation, a 
direct operational interpretation of the R\'enyi relative entropies was missing for a long 
time. This gap was filled in the classical case by Csisz\'ar \cite{Csiszar}, who defined the 
operational notion of \ki{cutoff rates} and showed that the $\alpha$-relative 
entropies arise as cutoff rates in state discrimination problems. In Section \ref{sec:cutoff rates} we follow Csisz\'ar's approach to show that the $\alpha$-relative 
entropies can be given the same operational interpretation in the quantum case, at least for the parameter range $\alpha\in(0,1)$.

Given a state shared by several parties, and a statistical distance $D$,
the $D$-distance of the state from the set of uncorrelated states yields a measure of 
correlations among the parties. For instance, a popular measure of quantum correlations is 
the \ki{relative entropy of entanglement} \cite{VPRK}, which is the relative entropy distance of a 
multipartite quantum state from the set of separable (i.e., only classically correlated) 
states.
Similarly, a measure of the total amount of correlations between parties 
$A$ and $B$ sharing a bipartite quantum state $\rho_{AB}$, can be defined by 
the $D$-distance of $\rho_{AB}$ from the set of product states,
\begin{equation*}
I_D(A:B\,|\,\rho_{AB}):=\inf_{\sigma_A\in\S(\hil_A),\sigma_B\in\S(\hil_B)}\dist{\rho_{AB}}{\sigma_A
\otimes\sigma_B},
\end{equation*} 
where $\S(\hil_A)$ and $\S(\hil_B)$ denote the state spaces of parties $A$ and $B$, 
respectively. When the 
statistical distance is the relative entropy $S$, 
there is a unique product state closest to $\rho_{AB}$, which  
is the product $\rho_A\otimes\rho_B$ of the marginals of $\rho_{AB}$, and we have the 
identities
\begin{align}
I_S(A:B\,|\,\rho_{AB})&=\sr{\rho_{AB}}{\rho_A\otimes\rho_B}\nonumber\\
&=\inf_{\sigma_A\in\S(\hil_A)}\sr{\rho_{AB}}{\sigma_A\otimes\rho_B}\nonumber\\
&=\inf_{\sigma_B\in\S(\hil_B)}\sr{\rho_{AB}}{\rho_A\otimes\sigma_B}.\label{correlation}
\end{align}
These identities, however, are not valid any longer if $S$ is replaced with some other 
statistical distance $D$, and one may wonder which formula gives the ``right'' measure of correlations, i.e., which one admits an operational interpretation. 
When $D$ is an $\alpha$-relative entropy or a Hoeffding distance, an operational 
interpretation can be obtained for $D(\rho_{AB}\,||\,\rho_A\otimes\rho_B)$ in the setting of discriminating $\rho_{AB}$ from $\rho_A\otimes\rho_B$, as described above. It seems, however, that when $D$ is an $\alpha$-relative 
entropy and the aim is to measure correlations between the input and the output of a 
stochastic communication channel then 
it is the last formula in \eqref{correlation} (with $S$ replaced with an $\alpha$-relative entropy) that yields a natural operational interpretation, as we will see below.

By a \ki{classical-quantum communication channel} (or simply a channel) we mean a map 
$\ch:\,\X\to\S(\hil)$, where $\X$ is a set and $\hil$ is a Hilbert space, which we
assume to be finite-dimensional. 
Note that there is no restriction on the cardinality of $\X$, and this formulation encompasses both the case of classical channels (i.e., when the range of $W$ is commutative) and the standard formalism for quantum channels (i.e., when $\X$
is the state space of an input 
Hilbert space and $W$ is a completely positive trace-preserving map).
A ``lifting'' of the 
channel can be defined by 
$\hat\ch:\,\X\to\S(\hil_\X\otimes \hil),\ds
 \hat\ch:\,x\mapsto \delta_x\otimes W_x$,
where $\hil_\X$ is some auxiliary Hilbert space with dimension equal to the cardinality of $\X$, and $\delta_x:=\pr{e_x}$ for some orthonormal system $\{e_x\}_{x\in\X}$ in $\hil_\X$.
The expectation value of $\hat\ch$ with respect to a finitely supported probability measure
$p\in\M_f(\X)$ is a classical-quantum state $\Exp_p\hat\ch=\sum_x p(x)\delta_x\otimes W_x
$
on the joint system of the input and the output of the channel, and its marginals are given by $\Tr_{\hil}\Exp_p\hat\ch=\hat p:=\sum_x p(x)\delta_x$ and 
$\Tr_{l^2(\X)} \Exp_p\hat\ch=\Exp_p\ch=\sum_x p(x)W_x$.
The amount of correlations between the input and the output in the state $\Exp_p\hat\ch$, as measured by the relative entropy, can be written in various equivalent ways:
\begin{align}
&I_S(p;W)\nonumber\\
&:=\sr{\Exp_p\hat\ch}{\hat p\otimes \Exp_p W}
=\inf_{\sigma\in S(\hil)}\sr{\Exp_p\hat\ch}{\hat p\otimes \sigma}\label{classical-quantum correlations1}\\
&=
\sum_x p(x)\sr{W_x}{\Exp_p W}
=
\inf_{\sigma\in\S(\hil)}\sum_x p(x)\sr{W_x}{\sigma}\label{classical-quantum correlations2}\\
&=
S(\Exp_pW)-\sum_x p(x)S(W_x).\label{classical-quantum correlations}
\end{align}
The Holevo-Schumacher-Westmoreland theorem \cite{Holevo3,SW} shows that the asymptotic information transmission capacity of a channel, under the assumption of product encoding, is given by 
the \ki{Holevo capacity} 
\begin{equation}\label{Holevo capacity}
\chi_{S}^*(W):=\sup_{p\in\M_f(\X)}I_S(p;W),
\end{equation}
which is the maximal amount of correlation that can be created between the classical input 
and the quantum output in a classical-quantum state of the form $\Exp_p \hat 
W,\,p\in\M_f(W)$. A geometric interpretation of the Holevo capacity was given in 
\cite{OPW}, where it was shown that the Holevo capacity of a channel $W$ is equal to the relative entropy radius $R_S(\ran W)$ of its range, where the $D$-radius of a subset $\Sigma\subset\S(\hil)$ for a statistical distance $D$ is defined as
 \begin{equation}\label{divergence radius}
 R_D(\Sigma):=\inf_{\sigma\in\S(\hil)}\sup_{\rho\in\Sigma}D(\rho\,||\,\sigma).
 \end{equation}
Not so suprisingly, the identities in \eqref{classical-quantum correlations1}--\eqref{classical-quantum correlations} do not hold for a general statistical distance $D$, and one may define various formal generalizations of the Holevo capacity. Here we will be interested in the quantities
\begin{align}
\chi^*_{D,0}(\ch)&:=\sup_{p\in\M_f(\X)} D(\Exp_p\hat\ch\,||\,\D{p}\otimes\Exp_p\ch),\label{def:cap0}\\
\chi^*_{D,1}(\ch)&:=\sup_{p\in\M_f(\X)}\inf_{\sigma\in\S(\hil)} D(\Exp_p\hat\ch\,||\,\D{p}\otimes\sigma),\label{def:cap1}\\
\chi^*_{D,2}(\ch)&:=\sup_{p\in\M_f(\X)}\inf_{\sigma\in\S(\hil)}\sum_{x\in\X} p(x)D(W_x\,||\,\sigma),\label{def:cap2}\\
R_{D}(\ran\ch)&:=\inf_{\sigma\in\S(\hil)}\sup_{x\in\X} D(W_x\,||\,\sigma).\label{def:cap3}
\end{align}
The capacities $\chi^*_{D,1}(\ch),\chi^*_{D,2}(\ch)$ and $R_{D}(\ran\ch)$ were shown to be equal in \cite{Csiszar} when the channel is classical and $D$ is an $\alpha$-relative entropy $S_\alpha$ with arbitrary non-negative parameter $\alpha$, and in \cite{KW}, the identity 
$\chi^*_{S_\alpha,1}(\ch)=R_{S_\alpha}(\ran\ch)$ was shown for quantum channels and $\alpha\in(1,+\infty)$. In Section \ref{sec:equivalence} we follow the approach of \cite{Csiszar} to show that $\chi^*_{D,1}(\ch)=\chi^*_{D,2}(\ch)=R_{D}(\ran\ch)$ for classical-quantum channels when $D$ is an $\alpha$-relative entropy with parameter $\alpha\in(0,2]$.

The Holevo-Schumacher-Westmoreland theorem identifies the Holevo capacity \eqref{Holevo capacity} as the optimal rate of information transmission through the channel in an asymptotic scenario, under the assumption that the noise described by the channel occurs independently at consecutive uses of the channel (memoryless channel).
However, in practical applications one can use a channel only finitely many times, and the memoryless condition might not always be realistic, either. Hence, it is desirable to have bounds on the information transmission capacity of a channel for finitely many uses. For a given threshold $\ep>0$, the \ki{one-shot $\ep$-capacity} of the channel is the maximal number of bits that can be transmitted by one single use of the channel, with an average error not exceeding $\ep$. Note that finitely many (possibly correlated) uses of a channel can be described as the action of one single channel acting on sequences of inputs, and hence the study of one-shot capacities addresses the generalization of coding theorems in the direction of finitely many uses and possibly correlated channels at the same time.
In \cite{MD} a lower bound on the one-shot $\ep$-capacity of an arbitrary classical-quantum channel $W$ was given in terms of the R\'enyi capacities $\chi^*_{S_\alpha,0}(\ch)$ with parameter $\alpha\in [0,1)$.
This bound was shown to be asymptotically optimal in the sense of yielding the Holevo 
capacity as a lower bound in the asymptotic limit, but no upper bound of similar form has 
been known up till now. In Section \ref{sec:one-shot capacities} we show an upper bound on 
the one-shot $\ep$-capacity in 
terms of the R\'enyi capacities $\chi^*_{S_\alpha,1}(\ch)$ with parameter $\alpha>1$ that is 
again asymptotically optimal in the above sense. It remains an open question whether the capacities $\chi^*_{S_\alpha,0}(\ch)$ and $\chi^*_{S_\alpha,1}(\ch)$ are equal for a given $\alpha$. To the best of our knowledge, the answer to this question is unknown even in the classical case.

\section{Preliminaries on the R\'enyi relative entropies}

Let $\hil$ be a finite-dimensional Hilbert space with $d:=\dim\hil$. We will use the notations
$\B(\hil)_+$ and $\B(\hil)_{++}$ to denote the positive semidefinite and the strictly positive definite operators on $\hil$, respectively. Similarly, we denote the set of density operators
(positive semidefinite operators with unit trace) by $\S(\hil)$, and use the notation
$\S(\hil)_{++}$ for the set of invertible density operators.
We will use the conventions $0^{\alpha}:=0,\s \alpha\in\R$, and $\log 0:=-\infty,\s \log+\infty:=+\infty$. By the former, powers of a positive semidefinite operator are only taken on its support, i.e., if the spectral decomposition of an $A\in\B(\hil)_+$ is $A=\sum_k a_k P_k$, where all $a_k>0$, then $A^\alpha:=\sum_k a_k^\alpha P_k$ for all $\alpha\in\R$.  
In particular, $A^0$ is the projection onto the support of $A$.

Following \cite{Petz}, we define for every $\alpha\in [0,+\infty)\setminus\{1\}$ the \ki{$\alpha$-quasi-relative entropy} of an $A\in\B(\hil)_+$ with respect to a $B\in\B(\hil)_+$ as
\begin{align*}
&\qrsr{A}{B}{\alpha}\\
&:=
\begin{cases}
\sgn(\alpha-1)\Tr A^\alpha B^{1-\alpha},& \supp A\le\supp B\\
&\text{or }\alpha\in[0,1),\\
+\infty,&\text{otherwise}.
\end{cases}
\end{align*}
The \ki{R\'enyi $\alpha$-relative entropy} of $A$ with respect to $B$ is then defined as
\begin{equation*}
\rsr{A}{B}{\alpha}:=\frac{1}{\alpha-1}\log\sgn(\alpha-1)\qrsr{A}{B}{\alpha}.
\end{equation*}
Note that $\rsr{A}{B}{\alpha}=+\infty$ if $\supp A\perp\supp B$, or if $\supp A\nleq\supp B$ and $\alpha>1$. In all other cases, $\rsr{A}{B}{\alpha}$ is a finite number, given by 
$\rsr{A}{B}{\alpha}=\frac{1}{\alpha-1}\log\Tr A^\alpha B^{1-\alpha}$.
Note that for $\alpha\in(0,1)$, we have
\begin{equation}\label{duality}
\rsr{A}{B}{1-\alpha}=\frac{1-\alpha}{\alpha}\rsr{B}{A}{\alpha}.
\end{equation}
It is easy to see that if $\Tr A=1$ then
\begin{equation*}
\rsr{A}{B}{1}:=\lim_{\alpha\to 1}\rsr{A}{B}{\alpha}=\sr{A}{B}
\end{equation*}
where $\sr{A}{B}$ is the \ki{relative entropy}
\begin{equation*}
\sr{A}{B}:=\begin{cases}
\Tr A(\log A-\log B),& \supp A\le\supp B,\\
+\infty,&\text{otherwise}.
\end{cases}
\end{equation*}

Operator monotonicity of the function $x\mapsto x^{1-\alpha},\,x\ge 0$, for 
$\alpha\in[0,1]$ yields that 
\begin{align*}
\qrsr{A}{B+C}{\alpha}&\le\qrsr{A}{B}{\alpha}\ds\text{and}\\
\rsr{A}{B+C}{\alpha}&\le\rsr{A}{B}{\alpha}
\end{align*}
for any $A,B,C\in \B(\hil)_+$ and $\alpha\in[0,1]$, and the same holds for $\alpha> 1$ if $B$ and $C$ commute. In 
particular, for fixed $A,B\in\B(\hil)_+$, the maps $0<\ep\mapsto \qrsr{A}{B+\ep I}{\alpha}$ and 
$0<\ep\mapsto \rsr{A}{B+\ep I}{\alpha}$ are monotonic decreasing, and it is easy to see that, 
for any $\alpha\in[0,+\infty)$, 
\begin{align}
\qrsr{A}{B}{\alpha}&=\sup_{\ep> 0}\qrsr{A}{B+\ep I}{\alpha},\label{ep limit Q}\\
\rsr{A}{B}{\alpha}&=\sup_{\ep> 0}\rsr{A}{B+\ep I}{\alpha}.
\label{ep limit S}
\end{align}

For $\alpha\in[0,2]\setminus\{1\}$, the $\alpha$-quasi-relative entropies have the monotonicity property \cite{HMH,Petz,TCR}
\begin{equation}\label{monotonicity of quasi entropies}
\qrsr{\map(A)}{\map(B)}{\alpha}\le \qrsr{A}{B}{\alpha},\ds A,B\in\B(\hil)_+,
\end{equation}
where $\map$ is any completely positive trace-preserving (CPTP) map on $\B(\hil)$. 
As a consequence, the $\alpha$-quasi-relative entropies are jointly convex in their arguments for $\alpha\in [0,2]\setminus\{1\}$:
\begin{equation}\label{convexity of quasi-entropies}
 \qrsr{\sum\nolimits_i p_iA_i}{\sum\nolimits_i p_iB_i}{\alpha}\le \sum \nolimits_i p_i\qrsr{A_i}{B_i}{\alpha},
\end{equation}
where $A_i,B_i\in\B(\hil)_+$, and $\{p_i\}$ is a finite probability distribution
\cite{Ando,Lieb,Petz}.

The monotonicity property \eqref{monotonicity of quasi entropies} of the $\alpha$-quasi-relative entropies yields that, for any CPTP map $\map$ on $\B(\hil)$ and $\alpha\in[0,2]$,
\begin{equation*}
\rsr{\map(A)}{\map(B)}{\alpha}\le \rsr{A}{B}{\alpha},\ds A,B\in\B(\hil)_+.
\end{equation*}
Convexity of the function $\frac{1}{\alpha-1}\log$ for $\alpha\in[0,1)$ yields, by \eqref{convexity of quasi-entropies}, that for $\alpha\in[0,1]$,
\begin{equation}\label{joint convexity of relative entropies}
\rsr{\sum\nolimits_i p_iA_i}{\sum\nolimits_i p_iB_i}{\alpha}\le \sum\nolimits_i p_i\rsr{A_i}{B_i}{\alpha}
\end{equation}
for any finite probability distribution $\{p_i\}$ and $A_i,B_i\in\B(\hil)_+$. Note that the joint convexity \eqref{convexity of quasi-entropies} of the $\alpha$-quasi-relative entropies for $\alpha\in(1,2]$ is not inherited by the corresponding R\'enyi relative entropies, as 
$\frac{1}{\alpha-1}\log$ is not convex for $\alpha>1$; for a counterexample, see e.g.~\cite{BBR}. Actually, the example of \cite{BBR} shows that the 
R\'enyi relative entropies are not even convex in their first argument for $\alpha>1$.
However, we have the following:
\begin{thm}\label{thm:convexity}
For a fixed $A\in\B(\hil)_+$, the map $B\mapsto \rsr{A}{B}{\alpha}$ is convex on $\B(\hil)_+$ for every $\alpha\in[0,2]$.
\end{thm}
\begin{IEEEproof}
For $\alpha\in[0,1]$, the assertion is a weaker version of \eqref{joint convexity of relative entropies}, and hence for the rest we assume that $\alpha\in(1,2]$. Let $A,B_1,B_2\in\B(\hil)_+$; it suffices to show that 
\begin{align}
&\rsr{A}{\eta (B_1+\ep I)+(1-\eta)(B_2+\ep I)}{\alpha}\nonumber\\
&\ds\le \eta\rsr{A}{B_1+\ep I}{\alpha}+(1-\eta)\rsr{A}{B_2+\ep I}{\alpha}\label{convexity inequality}
\end{align}
holds for every $\eta\in(0,1)$. Taking the limit $\ep\searrow 0$ will then give the desired convexity inequality. Note that 
\eqref{convexity inequality} is equivalent to
\begin{align*}
&\log\omega(\eta (B_1+\ep I)+(1-\eta)(B_2+\ep I)^{1-\alpha})\\
&\ds\le \eta\log\omega((B_1+\ep I)^{1-\alpha})+(1-\eta)\log\omega((B_2+\ep I)^{1-\alpha}),
\end{align*}
where $\omega(X):=\Tr A^\alpha X,\,X\in\B(\hil)$, is a positive linear functional on $\B(\hil)$.
Proposition 1.1 in \cite{AH} states that the functional 
$X\mapsto \log\omega(f(X)),\,X\in\B(\hil)_{++}$, is convex whenever $\omega$ is a positive linear functional and $f$ is a non-negative operator monotone decreasing function on $(0,+\infty)$. Applying this to the $\omega$ above and $f(x):=x^{1-\alpha},\,x>0$, the assertion follows.
\end{IEEEproof}

By computing its second derivative, it is easy to see that the function
$\alpha\mapsto\log\Tr A^\alpha B^{1-\alpha},\, \alpha\in\R$,
is convex on $\R$ for any fixed $A,B\in\B(\hil)_+$, which yields by a simple computation the following:
\begin{lemma}\label{lemma:Renyi monotonicity}
If $\Tr A\le 1$ then the function $\alpha\mapsto \rsr{A}{B}{\alpha}$ is monotonically increasing on $[0,1)$ and on $(1,+\infty)$. Moreover, if $\Tr A=1$ then $\alpha\mapsto \rsr{A}{B}{\alpha}$ is monotonically increasing on $[0,+\infty)$.
\end{lemma}

\begin{prop}
Assume that $\Tr A\le 1$ and $\Tr B\le 1$. For $\alpha\in (0,1)$,
$\rsr{A}{B}{\alpha}\ge 0$ with equality
if and  only if $A=B$ and $\Tr A=1$. 
If $A$ is a density operator and $\Tr B\le 1$ then, for all $\alpha\in [1,+\infty)$,
$\rsr{A}{B}{\alpha}\ge 0$, and
$\rsr{A}{B}{\alpha}= 0$ if and  only if $A=B$.
Moreover, if both $A$ and $B$ are density operators then 
the Csisz\'ar-Pinsker inequality
\begin{equation*}
\rsr{A}{B}{\alpha}\ge\half\norm{A-B}_1^2
\end{equation*}
holds for all $\alpha\ge 1$.
\end{prop}
\begin{IEEEproof}
 Assume first that $\alpha\in [0,1)$. Then, by H\"older's inequality,
 \begin{equation*}
 \Tr A^\alpha B^{1-\alpha}\le \bz\Tr A\jz^\alpha\bz\Tr B\jz^{1-\alpha}\le 1,
 \end{equation*}
 from which $\rsr{A}{B}{\alpha}=\frac{1}{\alpha-1}\log\Tr A^\alpha B^{1-\alpha}\ge 0$. 
Obviously, $\rsr{A}{B}{\alpha}=0$ if and only if $\Tr A^\alpha B^{1-\alpha}=1$. By the above, this is true if and only if $\Tr A=\Tr B=1$, and H\"older's inequality holds with equality. The latter condition yields that $B=\lambda A$ for some $\lambda\ge 0$, and $\Tr A=\Tr B$ yields $\lambda=1$.
Lemma \ref{lemma:Renyi monotonicity} yields the assertion on strict positivity for $\alpha\ge 1$ when $A$ is a density operator. The Csisz\'ar-Pinsker inequality holds for $\alpha=1$ (cf.~Theorem 3.1 in \cite{HOT}) and hence, by Lemma \ref{lemma:Renyi monotonicity}, for all $\alpha\ge 1$.
\end{IEEEproof}

For a density operator $\rho\in\S(\hil)$, its \ki{R\'enyi $\alpha$-entropy} for $\alpha\in[0,+\infty)$ is
\begin{equation*}
S_\alpha(\rho):=\log d-\rsr{\rho}{(1/d)I}{\alpha}.
\end{equation*}
For $\alpha\ne 1$ we have $S_\alpha(\rho)=\frac{1}{1-\alpha}\log\Tr\rho^\alpha$, which is easily seen to be non-negative, and $\rsr{\rho}{(1/d)I}{\alpha}\ge 0$ yields that
\begin{equation}\label{Renyi entropy}
0\le S_\alpha(\rho)\le\log d,\ds\ds\ds \alpha\in[0,+\infty).
\end{equation}

The \ki{Hoeffding distance} of states $\rho,\sigma\in\S(\hil)$ with parameter $r\ge 0$ is defined as
\begin{align}
\hdist{\rho}{\sigma}{r}&:=\sup_{0\le \alpha<1}\left\{\frac{-\alpha r}{1-\alpha}+\rsr{\rho}{\sigma}{\alpha}\right\}\nonumber\\
&=\sup_{0\le \alpha<1}\frac{-\alpha r-\psi(\alpha)}{1-\alpha}=\sup_{s\ge 0}\{-sr-\tilde\psi(s)\},
\label{hdist def}
\end{align}
where 
\begin{align}
\psi(\alpha)&:=\log\Tr\rho^{\alpha}\sigma^{1-\alpha},\ds\alpha\in\R,\nonumber\\
\tilde\psi(s)&:=(1+s)\psi\bz s/(1+s)\jz,\ds s>-1.\label{psi tilde}
\end{align}
Convexity of $\psi$ yields the convexity of $\tilde\psi$, and a simple computation shows that
$\psi(0)+\psi'(0)=\tilde\psi'(0)\le\lim_{s\to\infty}\tilde\psi'(s)=\psi(1)\le 0$.
Hence, 
\begin{equation*}
\hdist{\rho}{\sigma}{r}=\begin{cases}
-\tilde\psi(0)=-\psi(0),& -r\le\psi(0)+\psi'(0),\\
+\infty,& -r>\psi(1).
\end{cases}
\end{equation*}

The function $r\mapsto\hdist{\rho}{\sigma}{r}$ is the Legendre-Fenchel transform (up to the sign of the variable) of $\tilde\psi$ on $[0,+\infty)$ and hence it is convex on $[0,+\infty)$. Using the bipolar theorem for convex functions \cite[Proposition 4.1]{ET}, we get
\begin{equation*}
\rsr{\rho}{\sigma}{\alpha}=-\sup_{r\ge 0}\left\{\frac{-r\alpha}{1-\alpha}-\hdist{\rho}{\sigma}{r}\right\},\ds\ds 0\le \alpha<1.
\end{equation*}
That is, the R\'enyi relative entropies with parameter in $[0,1)$ and the Hoeffding distances with parameter $r\ge 0$ mutually determine each other. Note that $r\mapsto \hdist{\rho}{\sigma}{r}$ is monotonic decreasing, and
\begin{equation*}
\rsr{\rho}{\sigma}{0}=\lim_{r\to\infty}\hdist{\rho}{\sigma}{r}\le\hdist{\rho}{\sigma}{0}=\rsr{\rho}{\sigma}{1}.
\end{equation*}

Finally, the \ki{max-relative entropy} of $A,B\in\S(\hil)_+$ was defined in \cite{Datta} as 
  $\srmax{A}{B}:=\inf\{\gamma\,:\,A\le 2^\gamma B\}$. One can easily see that if $A$ and $B$ 
 commute then $\srmax{A}{B}=\rsr{A}{B}{\infty}:=\lim_{\alpha\to\infty}\rsr{A}{B}{\alpha}$, but for non-commuting $A$ and $B$, $\srmax{A}{B}<\rsr{A}{B}{\infty}$ might happen \cite{MD}. In general, $\rsr{A}{B}{2}\le\srmax{A}{B}\le \rsr{A}{B}{\infty}$ \cite{BD,Colbeck}.

\section{Cutoff rates for quantum state discrimination}\label{sec:cutoff rates}

Consider the asymptotic binary state discrimination problem with null hypothesis $\rho$ and alternative hypothesis $\sigma$, as described in the Introduction. 
We will consider the scenario where the error probability of the 
second kind is minimized under an exponential constraint on the error probability of the 
first kind; the quantity of interest in this case is
\begin{align*}
\beta_{n,r}:=\min\{&\beta_n(T)\,|\,T\in\B(\hil^{\otimes n}),\,0\le T\le I,\\
&\text{and}\s\alpha_n(T)\le 2^{-nr}\},
\end{align*}
where $r$ is some fixed positive number. In general, there is no closed formula to express $\beta_{n,r}$ or the optimal measurement in terms of $\rho$ and $\sigma$ for a finite $n$, but it becomes possible in the limit of large $n$. We define the \ki{Hoeffding exponents} for a parameter $r>0$ as
\begin{align*}
\hli{\rho}{\sigma}{r}:=\inf_{\{T_n\}}\big\{&\liminf_{n\to\infty}
(1/n)\log\beta_{n}(T_n)\,\big|\,\\
&\limsup_{n\to\infty}(1/n)\log\alpha_n(T_n)< -r\big\},\\
\hls{\rho}{\sigma}{r}:=\inf_{\{T_n\}}\big\{&\limsup_{n\to\infty}
(1/n)\log\beta_{n}(T_n)\,\big|\,\\
&\limsup_{n\to\infty}(1/n)\log\alpha_n(T_n)< -r\big\},\\
\hl{\rho}{\sigma}{r}:=\inf_{\{T_n\}}\big\{&\lim_{n\to\infty}
(1/n)\log\beta_{n}(T_n)\,\big|\,\\
&\limsup_{n\to\infty}(1/n)\log\alpha_n(T_n)< -r\big\}.
\end{align*}
It is easy to see that 
\begin{align*}
\hli{\rho}{\sigma}{r}&\le\liminf_{n\to\infty}\frac{1}{n}\log\beta_{n,r}\\
&\le
\limsup_{n\to\infty}\frac{1}{n}\log\beta_{n,r}\le
\hls{\rho}{\sigma}{r}.
\end{align*}
Moreover, as it was shown in \cite{ANSzV,Hayashi,HMO2,Nagaoka}, we have
\begin{equation}\label{hoeffding theorem}
\hli{\rho}{\sigma}{r}=
\hls{\rho}{\sigma}{r}=\hl{\rho}{\sigma}{r}=-\hdist{\rho}{\sigma}{r},
\end{equation}
where $\hdist{\rho}{\sigma}{r}$ is the Hoeffding distance defined in \eqref{hdist def},
and hence, the limit $\lim_{n\to\infty}\frac{1}{n}\log\beta_{n,r}$ exists and
\begin{align*}
\lim_{n\to\infty}\frac{1}{n}\log\beta_{n,r}=-\hdist{\rho}{\sigma}{r}.
\end{align*}

Note that while the above result gives the exact value of the optimal exponential decay rate for every $r$, the evaluation of $\hdist{\rho}{\sigma}{r}$ is a non-trivial task even for one single $r$. Indeed, there is no closed formula known for the Hoeffding distance in general, and, as the definition \eqref{hdist def} shows, in order to compute $\hdist{\rho}{\sigma}{r}$, one has to know in principle all the R\'enyi relative entropies $\rsr{\rho}{\sigma}{\alpha}$ for every $\alpha\in(0,1)$, and solve an optimization problem. It is thus natural to look for simple approximants of the function $r\mapsto \hdist{\rho}{\sigma}{r}$ for given $\rho$ and $\sigma$.
Following \cite{Csiszar}, for a $\kappa<0$ we define the \ki{generalized $\kappa$-cutoff rate} $\cutoff{\rho}{\sigma}{\kappa}$ as the supremum of all $r_0\ge 0$ that satisfy
\begin{equation}\label{def:cutoff rate}
\hls{\rho}{\sigma}{r}\le\kappa(r_0-r),\ds\ds\ds r\ge 0.
\end{equation} 
That is, we are looking for a linear approximation of $r\mapsto \hdist{\rho}{\sigma}{r}$ which is optimal among all the linear functions with a given slope.
Note that \eqref{def:cutoff rate} gives a restriction only for $r\le r_0$, as otherwise the right-hand side is non-negative and the inequality holds trivially.
That is, one can ensure an exponential decay rate at least as fast as given in the right-hand side of \eqref{def:cutoff rate} whenever $r<r_0:=\cutoff{\rho}{\sigma}{\kappa}$. 
Moreover, as the following Theorem shows, the cutoff rate is easy to evaluate, as it is equal to a R\'enyi relative entropy with a given parameter depending on $\kappa$.

\begin{thm}\label{thm:cutoff}
For every $\kappa<0$,
\begin{equation}\label{cutoff rates}
\cutoff{\rho}{\sigma}{\kappa}=\frac{1}{|\kappa|}\rsr{\rho}{\sigma}{\frac{|\kappa|}{1+|\kappa|}}=\rsr{\sigma}{\rho}{\frac{1}{1+|\kappa|}}.
\end{equation}
\end{thm}
\begin{IEEEproof}
If $\supp\rho\perp\supp\sigma$ then all the quantities in \eqref{cutoff rates} are $+\infty$ and the assertion holds trivially. Hence, for the rest we assume that $\supp\rho$ is not orthogonal to $\supp\sigma$.
Note that the second identity follows from \eqref{duality}. 
Let $\kappa<0$ be fixed.
By \eqref{hoeffding theorem}, our goal is to determine the largest $r_0$ such that 
\begin{equation*}
-|\kappa|r+|\kappa|r_0\le -\hls{\rho}{\sigma}{r}=\hdist{\rho}{\sigma}{r},\ds\ds\ds r\ge 0.
\end{equation*}
By \eqref{hdist def}, $\hdist{\rho}{\sigma}{r}\ge-|\kappa|r-\tilde\psi(|\kappa|)$ for every $r\ge 0$, where $\tilde\psi$ is given in \eqref{psi tilde}. On the other hand, 
for $r_\kappa:=-\tilde\psi'(|\kappa|)$ we have $\tilde\psi(s)\ge \tilde\psi(|\kappa|)+(s-|\kappa|)\tilde\psi'(|\kappa|),\,s\ge 0$, due to the convexity of $\tilde\psi$ and hence,
\begin{align*}
\hdist{\rho}{\sigma}{r_\kappa}&=
\sup_{s\ge 0}\{s\tilde\psi'(|\kappa|)-\tilde\psi(s)\}
=|\kappa|\tilde\psi'(|\kappa|)-\tilde\psi(|\kappa|)\\
&=-|\kappa|r_\kappa-\tilde\psi(|\kappa|).
\end{align*} 
Therefore,
\begin{align*}
\cutoff{\rho}{\sigma}{\kappa}&=-\frac{1}{|\kappa|}\tilde\psi(|\kappa|)=-\frac{1+|\kappa|}{|\kappa|}\psi\bz\frac{|\kappa|}{1+|\kappa|}\jz\\
&=\frac{1}{|\kappa|}\rsr{\rho}{\sigma}{\frac{|\kappa|}{1+|\kappa|}}.\qedhere
\end{align*}
\end{IEEEproof}

The following Corollary is immediate from Theorem \ref{thm:cutoff}, and gives an operational interpretation of the R\'enyi relative entropies with parameter between $0$ and $1$:
\begin{cor}
For every $\rho,\sigma\in\S(\hil)$ and every $\alpha\in(0,1)$,
\begin{equation*}
\rsr{\rho}{\sigma}{\alpha}=\frac{\alpha}{1-\alpha}\cutoff{\rho}{\sigma}{\frac{\alpha}{\alpha-1}}=\cutoff{\sigma}{\rho}{\frac{\alpha-1}{\alpha}}.
\end{equation*}
\end{cor}
\medskip

In the above, we considered the scenario where the consecutive trials are independent and 
identically distributed, and hence the state describing the outcome probabilities of $n$ 
trials is a state of the form $\rho^{\otimes n}$ or $\sigma^{\otimes n}$.
In a more general scenario, that encompasses correlated trials, one can consider a sequence 
of Hilbert spaces $\vec{\hil}:=\{\hil_n\}_{n\in\N}$ and two sequences of states 
$\vec{\rho}:=\{\rho_n\}_{n\in\N}$ and $\vec{\sigma}:=\{\sigma_n\}_{n\in\N}$. The goal is 
again to analyze the asymptotic performance of a decision scheme for deciding between 
$\rho_n$ and $\sigma_n$ for each $n\in\N$. The error probabilities $\alpha_n$ and $\beta_n$ 
can be defined in the same way as above, and in analogy with the above problem, the limit 
$\lim_{n\to\infty}(1/c(n))\log\beta_{n,r}$ can be considered, where $c:\,\N\to\N$ is some 
monotonically increasing function such that $\lim_{n\to\infty}c(n)=+\infty$. The following 
was shown in \cite{HMO2}:
\begin{thm}\label{thm:HMO2}
Assume that the limit 
$\psi(\alpha):=\lim_{n\to\infty}\frac{1}{c(n)}(\alpha-1)\rsr{\rho_n}{\sigma_n}{\alpha}$ 
exists for all $\alpha\in[0,1)$ and the convergence is uniform on $[0,1)$. Assume, moreover, 
that $\psi$ is differentiable on $(0,1)$. Then,
\begin{align*}
\lim_{n\to\infty}\frac{1}{c(n)}\log\beta_{n,r}&=-\lim_{n\to\infty}\frac{1}{c(n)}\hdist{\rho_n}{\sigma_n}{c(n)r}\\
&=:-\hdist{\vec{\rho}}{\vec{\sigma}}{r}.
\end{align*}
Moreover, $\hdist{\vec{\rho}}{\vec{\sigma}}{r}=
\sup_{0\le\alpha<1}\left\{\frac{-\alpha r}{1-\alpha}+\frac{\psi(\alpha)}{\alpha-1}\right\}$, where
\begin{equation*}
\frac{\psi(\alpha)}{\alpha-1}=
\rsr{\vec{\rho}}{\vec{\sigma}}{\alpha}:=\lim_{n\to\infty}\frac{1}{c(n)}\rsr{\rho_n}{\sigma_n}{\alpha}.
\end{equation*}
\end{thm}
\smallskip

A particular example that satisfies the conditions of Theorem \ref{thm:HMO2} is the case 
where $\rho_n$ and $\sigma_n$ are the $n$-step restrictions of classical ergodic Markov 
chains with finite state-space \cite{HMO2}. Physically motivated examples can be obtained by 
considering $\rho_n$ and $\sigma_n$ to be finite-block restrictions of temperature states of 
non-interacting fermionic and bosonic systems on cubic lattices \cite{MHOF,M}.

The cutoff rates $\cutoff{\vec{\rho}}{\vec{\sigma}}{\kappa}$ can again be defined in the same 
way as in \eqref{def:cutoff rate} (with the scale $1/n$ replaced with $1/c(n)$ in the 
definition of $\hls{\vec{\rho}}{\vec{\sigma}}{r}$). The same argument as in the proof of Theorem \ref{thm:cutoff} leads to the following:
\begin{thm}
Under the assumptions of Theorem \ref{thm:HMO2}, we have
\begin{equation*}
\cutoff{\vec{\rho}}{\vec{\sigma}}{\kappa}=\frac{1}{|\kappa|}\rsr{\vec{\rho}}{\vec{\sigma}}{\frac{|\kappa|}{1+|\kappa|}}=\rsr{\vec{\sigma}}{\vec{\rho}}{\frac{1}{1+|\kappa|}}
\end{equation*}
for every $\kappa<0$, or equivalently, for every $\alpha\in(0,1)$,
\begin{equation*}
\rsr{\vec{\rho}}{\vec{\sigma}}{\alpha}=\frac{\alpha}{1-\alpha}\cutoff{\vec{\rho}}{\vec{\sigma}}{\frac{\alpha}{\alpha-1}}=\cutoff{\vec{\sigma}}{\vec{\rho}}{\frac{\alpha-1}{\alpha}}.
\end{equation*}
\end{thm}

\section{Equivalence of capacities}\label{sec:equivalence}

Let $W:\,\X\to\S(\hil)$ be a classical-quantum channel as in the Introduction. Our aim in 
this section is to show that the capacities defined in \eqref{def:cap1}--\eqref{def:cap3} 
are equal to each other when $D=S_\alpha$ is a R\'enyi relative entropy with parameter 
$\alpha\in(0,2]$. We will assume that $\ran\ch$ is compact in $\S(\hil)$. This assumption is 
satisfied when $W$ is a CPTP map on the state space of an input Hilbert space as well as 
when $\X$ is a finite set.

Note that $\cS(\cH)$ is a compact convex subset of the Euclidean space $B(\cH)_{sa}$ (with
the Hilbert-Schmidt norm). Let $\cK$ be a compact subset of $\cS(\cH)$ and $\M(\cK)$ be
the set of all Borel probability measures on $\cK$. Let $C_\bR(\cK)$ be the real Banach
space of all real continuous functions on $\cK$ with the sup-norm; then $\M(\cK)$ is
identified with a w*-compact convex subset of the dual Banach space $C_\bR(\cK)^*$.
We also introduce the subset $\M_f(\cK)$ of $\M(\cK)$, consisting of finitely
supported measures.

For every $\alpha\in(0,2]\setminus\{1\}$ and $\eps\ge 0$, define the functions
$f_{\alpha,\eps}$ and $g_{\alpha,\eps}$ on $\M(\K)\times\S(\hil)$ by
\begin{align*}
f_{\alpha,\eps}(p,\sigma)&:=\int_\cK S_\alpha(\rho\|\sigma+\eps I)\,dp(\rho),\\
g_{\alpha,\eps}(p,\sigma)&:=\int_\cK Q_\alpha(\rho\|\sigma+\eps I)\,dp(\rho).
\end{align*}
Note that for every fixed $\sigma$, the functions 
$\rsr{\cdot}{\sigma+\ep I}{\alpha}$ and $\qrsr{\cdot}{\sigma+\ep I}{\alpha}$ are continuous for $\ep>0$ and, by \eqref{ep limit Q} and \eqref{ep limit S}, are lower semicontinuous for $\ep=0$. Hence, the integrals defining 
$f_{\alpha,\ep}$ and $g_{\alpha,\ep}$ exist for all $\ep\ge 0$.
Furthermore, by \eqref{ep limit Q}, \eqref{ep limit S}, and Beppo Levi's theorem,
\begin{equation}\label{f_alpha}
f_{\alpha,0}(p,\sigma)=\lim_{\eps\searrow0}f_{\alpha,\eps}(p,\sigma)
=\sup_{\eps>0}f_{\alpha,\eps}(p,\sigma),\qquad p\in\M(\cK),
\end{equation}
and the same holds if we replace $f_{\alpha,0}$ with $g_{\alpha,0}$ and $f_{\alpha,\ep}$ with $g_{\alpha,\ep}$.

\begin{lemma}\label{lemma:affine cont}
For every $\sigma\in\cS(\cH)$ and $\ep>0$, $f_{\alpha,\eps}(\cdot,\sigma)$ and $g_{\alpha,\eps}(\cdot,\sigma)$ are affine and continuous
on $\M(\cK)$.
\end{lemma}
\begin{IEEEproof}
The claims about the affinity are obvious, and
the continuity of the functions $\rsr{\cdot}{\sigma+\ep I}{\alpha}$ and 
$\qrsr{\cdot}{\sigma+\ep I}{\alpha}$ yields, by 
definition, that $f_{\alpha,\eps}(\cdot,\sigma)$ and $g_{\alpha,\eps}(\cdot,\sigma)$ are continuous in the w$^*$-topology.
\end{IEEEproof}

\begin{lemma}\label{lemma:convex cont}
For every $p\in\M(\cK)$ and $\ep>0$, $f_{\alpha,\eps}(p,\cdot)$ and $g_{\alpha,\eps}(p,\cdot)$ are convex and continuous on
$\cS(\cH)$.
\end{lemma}
\begin{IEEEproof}
Convexity follows from Theorem \ref{thm:convexity} and \eqref{convexity of quasi-entropies}. Let $\{\sigma_k\}_{k\in\N}$ be a sequence in $\S(\hil)$, converging to some $\sigma_0\in\S(\hil)$. 
Let $f_k(\rho):=\Tr\rho^\alpha(\sigma_k+\eps I)^{1-\alpha}$ and
$f(\rho):=\Tr\rho^\alpha(\sigma_0+\ep I)^{1-\alpha},\,\rho\in\K$. Since
\begin{align*}
&|\Tr\rho^\alpha(\sigma_{k}+\eps I)^{1-\alpha}
-\Tr\rho^\alpha(\sigma_{0}+\eps I)^{1-\alpha}|\\
&\ds\le\Tr\rho^\alpha\cdot\|(\sigma_{k}+\eps I)^{1-\alpha}
-(\sigma_0+\eps I)^{1-\alpha}\|_\infty,
\end{align*}
and $\Tr\rho^\alpha\le d$ for every $\alpha\ge 0$, 
we see that $\lim_k f_k(\rho)=f(\rho)$ uniformly in $\rho$. This yields the continuity 
of $g_{\alpha,\eps}(p,\cdot)$.

For $\alpha\in(1,2]$, 
$
f(\rho)\ge \Tr\rho^\alpha(1+\ep)^{1-\alpha}\ge(1+\ep)^{1-\alpha}d^{1-\alpha}$, due to \eqref{Renyi entropy}.
For $\alpha\in(0,1)$, the operator monotonicity of the function $x\mapsto x^{1-\alpha},\,x\ge 0$,  yields that $f(\rho)\ge \Tr\rho^{\alpha}(\ep I)^{1-\alpha}\ge\ep^{1-\alpha}$ for all $\rho\in\K$. Since
\begin{equation*}
|f_k(\rho)-f(\rho)|=f(\rho)\left|\frac{f_k(\rho)}{f(\rho)}-1\right|\ge\inf_{\rho\in\K}f(\rho)\left|\frac{f_k(\rho)}{f(\rho)}-1\right|,
\end{equation*}
we see that $f_k(\rho)/f(\rho)$ converges to $1$ uniformly in $\rho$ as $k\to\infty$, and 
hence 
\begin{equation*}
S_\alpha(\rho\|\sigma_{k}+\eps I)-S_\alpha(\rho\|\sigma_0+\ep I)=\frac{1}{\alpha-1}\log\frac{f_k(\rho)}{f(\rho)}
\end{equation*}
converges to $0$ uniformly in $\rho$, due to which $\lim_{k\to\infty}f_{\alpha,\ep}(p,\sigma_k)=f_{\alpha,\ep}(p,\sigma_0)$.
\end{IEEEproof}
\medskip

To simplify notation, we fix an $\alpha\in(0,2]\setminus\{1\}$ for the rest. We have the following:
\begin{prop}\label{prop:minimax}
For every $\ep>0$, there exists a $\sigma_\ep\in\S(\hil)$ such that 
\begin{align}
&\max_{p\in\M(\cK)}f_{\alpha,\eps}(p,\sigma_\eps)\nonumber\\
&\ds=\min_{\sigma\in\cS(\cH)}\max_{p\in\M(\cK)}f_{\alpha,\eps}(p,\sigma)
=\max_{p\in\M(\cK)}\min_{\sigma\in\cS(\cH)}f_{\alpha,\eps}(p,\sigma)\label{minimiax identity}\\
&\ds=
\min_{\sigma\in\S(\hil)}\max_{\rho\in\K}\rsr{\rho}{\sigma+\ep I}{\alpha}
=\max_{\rho\in\K}\rsr{\rho}{\sigma_\eps+\ep I}{\alpha}.\label{trivial identities}
 \end{align}
Moreover, the same relations hold if the maxima over $\M(\K)$ are replaced with maxima over $\M_f(\K)$.
\end{prop}
\begin{IEEEproof}
For a fixed $\sigma$, $f_{\alpha,\ep}(\cdot,\sigma)$ is continuous 
and, consequently, $p\mapsto\min_{\sigma\in\cS(\cH)}f_{\alpha,\eps}(p,\sigma)$ is upper semicontinuous and therefore they reach their suprema on the compact set $\M(\K)$.
Moreover, $f_{\alpha,\eps}(p,\sigma)\le \sup_{\rho\in\supp p}\rsr{\rho}{\sigma+\ep I}{\alpha},\,p\in\M(\K),\,\sigma\in\S(\hil)$, yields that the maximum of $f_{\alpha,\eps}(\cdot,\sigma)$ on $\M(\K)$ is reached at a Dirac probability measure and hence,
\begin{align}
\max_{p\in\M(\cK)}f_{\alpha,\eps}(p,\sigma)&=\max_{\rho\in\K}\rsr{\rho}{\sigma+\ep I}{\alpha}\nonumber\\
&=\max_{p\in\M_f(\cK)}f_{\alpha,\eps}(p,\sigma)\label{suprema identities}
\end{align}
for every $\sigma\in\S(\hil)$.
Continuity of $f_{\alpha,\ep}(p,\cdot)$ yields that $\sigma\mapsto \max_{p\in\M(\cK)}f_{\alpha,\eps}(p,\sigma)$ is lower semicontinuous on $\S(\hil)$ and hence it reaches its infimum at some point $\sigma_\ep$, which yields 
$\min_{\sigma\in\cS(\cH)}\max_{p\in\M(\cK)}f_{\alpha,\eps}(p,\sigma)=\max_{p\in\M(\cK)}f_{\alpha,\eps}(p,\sigma_\eps)$.
The identity of the two expressions in \eqref{minimiax identity} follows by Sion's minimax theorem \cite{Sion,Komiya}, due to Lemmas \ref{lemma:affine cont} and \ref{lemma:convex cont}.
The formulas in \eqref{trivial identities} follow from \eqref{suprema identities}.
The last assertion follows from \eqref{suprema identities} and the fact that
$f_{\alpha,\eps}|_{\M_f(\cK)\times\cS(\cH)}$ also satisfies the conditions in Sion's minimax theorem.
\end{IEEEproof}

For the rest, for every $\ep>0$ we fix a $\sigma_\ep$ as given in Proposition 
\ref{prop:minimax}. 
Note that the compactness of $\S(\hil)$ yields that
there exists a sequence $\{\ep_k\}_{k\in\N}$ and a $\sigma_0\in\S(\hil)$ such that 
$\lim_k\ep_k=0$ and $\lim_k\sigma_{\ep_k}=\sigma_0$. 

\begin{prop}\label{thm:minimax}
Let $\sigma_0$ be a limit point as above. Then,
\begin{align}
&\sup_{p\in\M(\cK)}f_{\alpha,0}(p,\sigma_0)\nonumber\\
&\ds=
\min_{\sigma\in\cS(\cH)}\sup_{p\in\M(\cK)}f_{\alpha,0}(p,\sigma)
=\sup_{p\in\M(\cK)}\min_{\sigma\in\cS(\cH)}f_{\alpha,0}(p,\sigma)\label{minimax identities 2}\\
&\ds=
\min_{\sigma\in\S(\hil)}\sup_{\rho\in\K}\rsr{\rho}{\sigma}{\alpha}
=\sup_{\rho\in\K}\rsr{\rho}{\sigma_0}{\alpha}.\label{trivial identities 2}
\end{align}
Moreover, the same relations hold if the suprema over $\M(\K)$ are replaced with suprema over $\M_f(\K)$.
\end{prop}
\begin{IEEEproof}
By \eqref{f_alpha}, $f_{\alpha,0}(p,\cdot)$ is lower semicontinuous on $\S(\hil)$ and hence so is the function $\sigma\mapsto\sup_{p\in\M(\cK)}f_{\alpha,0}(p,\sigma),\,\sigma\in\B(\hil)_+$. Therefore, they reach their infima on $\S(\hil)$.
For every $k\in\N$,
\begin{align}
\max_{p\in\M(\cK)}f_{\alpha,0}(p,\sigma_{\ep_k}+\ep_kI)&=\max_{p\in\M(\cK)}f_{\alpha,\ep_k}(p,\sigma_{\ep_k})\nonumber\\
&=\max_{p\in\M(\cK)}\min_{\sigma\in\cS(\cH)}f_{\alpha,\ep_k}(p,\sigma)\nonumber\\
&\le
\sup_{p\in\M(\cK)}\min_{\sigma\in\cS(\cH)}f_{\alpha,0}(p,\sigma),\label{epsilon minimax}
\end{align}
where the first identity is by definition, the second is due to Proposition \ref{prop:minimax}, and the inequality follows from \eqref{f_alpha}. Furthermore,
\begin{align*}
\sup_{p\in\M(\cK)}\min_{\sigma\in\cS(\cH)}f_{\alpha,0}(p,\sigma)
&\le\min_{\sigma\in\cS(\cH)}\sup_{p\in\M(\cK)}f_{\alpha,0}(p,\sigma)\\
&\le\sup_{p\in\M(\cK)}f_{\alpha,0}(p,\sigma_0)\\
&\le
\liminf_{k\to\infty}\sup_{p\in\M(\cK)}f_{\alpha,0}(p,\sigma_{\ep_k}+\ep_kI)\\
&\le
\sup_{p\in\M(\cK)}\min_{\sigma\in\cS(\cH)}f_{\alpha,0}(p,\sigma),
\end{align*}
where the first two inequalities are obvious, the third one follows from the
lower semicontinuity of 
$\sigma\mapsto\sup_{p\in\M(\cK)}f_{\alpha,0}(p,\sigma),\,\sigma\in\B(\hil)_+$, and the 
last inequality is due to \eqref{epsilon minimax}. This gives the identities in 
\eqref{minimax identities 2}, and the identities in \eqref{trivial identities 2} follow the 
same way as in Proposition \ref{prop:minimax}. The last assertion follows by repeating the argument above with the suprema and maxima over $\M(\K)$ replaced with suprema over $\M_f(\K)$.
\end{IEEEproof}

\begin{remark}
Note that the minima over $\S(\hil)$ in \eqref{minimax identities 2} and \eqref{trivial identities 2} can be replaced with infima over $\S(\hil)_{++}$.
\end{remark}
\begin{IEEEproof}
The trivial inequality $(1-\ep)\sigma+\ep(1/d)I\ge (1-\ep)\sigma$ yields
\begin{equation}\label{mixing inequality}
\rsr{\rho}{(1-\ep)\sigma+\ep(1/d)I}{\alpha}+\log(1-\ep)\le\rsr{\rho}{\sigma}{\alpha}
\end{equation}
for every $\ep\in(0,1)$, $\rho\in\K$ and $\sigma\in\B(\hil)$,
and hence, for every $p\in\M(\K)$,
\begin{equation}\label{mixing inequality2}
f_{\alpha,0}(p,(1-\ep)\sigma+\ep(1/d)I)+\log(1-\ep)\le f_{\alpha,0}(p,\sigma). \end{equation}
Thus, 
\begin{align*}
\inf_{\sigma\in\S(\hil)_{++}}f_{\alpha,0}(p,\sigma)\ge & \min_{\sigma\in\cS(\cH)}f_{\alpha,0}(p,\sigma)\\
\ge&
\min_{\sigma\in\S(\hil)}f_{\alpha,0}(p,(1-\ep)\sigma+\ep(1/d)I)\\
&+\log(1-\ep)\\
\ge&
\inf_{\sigma\in\S(\hil)_{++}}f_{\alpha,0}(p,\sigma)+\log(1-\ep),
\end{align*}
and by taking the supremum in $\ep$, we get $\inf_{\sigma\in\S(\hil)_{++}}f_{\alpha,0}(p,\sigma)= \min_{\sigma\in\cS(\cH)}f_{\alpha,0}(p,\sigma)$.
The assertion about the other two minima can be obtained by repeating the same argument after taking the supremum over $\rho\in\K$ in \eqref{mixing inequality} and the supremum over $p\in\M(\K)$ in \eqref{mixing inequality2}, respectively.
\end{IEEEproof}

\begin{remark}
The first supremum in \eqref{minimax identities 2} and the last one in \eqref{trivial identities 2} can be replaced with maxima.
\end{remark}
\begin{IEEEproof}
By Proposition \ref{thm:minimax},
\begin{align*}
\sup_{\rho\in\K}\rsr{\rho}{\sigma_0}{\alpha}&=\min_{\sigma\in\S(\hil)}\sup_{\rho\in\K}\rsr{\rho}{\sigma}{\alpha}\\
&\le
\sup_{\rho\in\K}\rsr{\rho}{(1/d)I}{\alpha}=
\sup_{\rho\in\K}\left\{ \log d-S_{\alpha}(\rho)\right\}\\
&\le \log d.
\end{align*}
Thus, 
$\rsr{\rho}{\sigma_0}{\alpha}$ is finite, and therefore it is given as
$\rsr{\rho}{\sigma_0}{\alpha}=\frac{1}{\alpha-1}\log\Tr\rho^{\alpha}\sigma_0^{1-\alpha}$ for every $\rho\in\K$. This yields that $\rho\mapsto\rsr{\rho}{\sigma_0}{\alpha}$ on $\K$ and $p\mapsto f_{\alpha,0}(p,\sigma_0)$ on $\M(\K)$ are continuous, and hence they reach their suprema.
\end{IEEEproof}

Since in the proofs of Propositions \ref{prop:minimax} and \ref{thm:minimax} we 
only used the properties of $f_{\alpha,\ep}$ established in Lemmas \ref{lemma:affine cont} 
and \ref{lemma:convex cont}, which are common with the properties of $g_{\alpha,\ep}$, we 
have the following:
\begin{prop}\label{prop:minimax3}
The assertions of Propositions \ref{prop:minimax} and \ref{thm:minimax} hold true if 
we replace $f_{\alpha,\ep}$ with $g_{\alpha,\ep}$ for all $\ep\ge 0$, and $S_\alpha$ with 
$Q_\alpha$.
\end{prop}

Now we are ready to prove the following:
\begin{thm}
Let $W:\,\X\to\S(\hil)$ be a classical-quantum channel with compact image. Then,
the capacities defined in \eqref{def:cap1}--\eqref{def:cap3} 
are equal to each other when $D=S_\alpha$ is a R\'enyi relative entropy with parameter 
$\alpha\in(0,2]$.
\end{thm}
\begin{IEEEproof}
The assertion is obvious for $\alpha=1$ from the identities \eqref{classical-quantum 
correlations1} and \eqref{classical-quantum correlations2}, so for the rest we assume that 
$\alpha\in(0,2]\setminus\{1\}$. Let $\K:=\ran\ch$. Proposition \ref{thm:minimax} yields that
\begin{align*}
\chi^*_{S_\alpha,2}(\ch)&=
\sup_{p\in\M_f(\X)}\inf_{\sigma\in\S(\hil)} \sum_x p(x)\rsr{W_x}{\sigma}{\alpha}\\
&=
\sup_{p\in\M_f(\K)}\inf_{\sigma\in\S(\hil)} \sum_{\rho\in\K} p(\rho)\rsr{\rho}{\sigma}{\alpha}\\
&=
\sup_{p\in\M_f(\K)}\min_{\sigma\in\S(\hil)}f_{\alpha,0}(p,\sigma)\\
&=
\min_{\sigma\in\S(\hil)}\sup_{\rho\in\K} \rsr{\rho}{\sigma}{\alpha}\\
&=
R_{S_\alpha}(\ran\ch).
\end{align*}
Let $\id$ be the identical channel on $\K=\ran W$, and let $\hat\id:\,\rho\mapsto\delta_\rho\otimes\rho$ be its lifting as in the 
Introduction. Using Proposition \ref{prop:minimax3}, we have
\begin{align*}
&\chi^*_{S_\alpha,1}(\ch)\\
&\ds=
\sup_{p\in\M_f(\X)}\inf_{\sigma\in\S(\hil)}\rsr{\Exp_p\hat\ch}{\D{p}\otimes\sigma}{\alpha}\\
&\ds=
\sup_{p\in\M_f(\K)}\inf_{\sigma\in\S(\hil)}\rsr{\Exp_p\hat\id}{\D{p}\otimes\sigma}{\alpha}\\
&\ds=
\sup_{p\in\M_f(\K)}\inf_{\sigma\in\S(\hil)}
\frac{1}{\alpha-1}\log\sgn(\alpha-1)g_{\alpha,0}(p,\sigma)\\
&\ds=
\frac{1}{\alpha-1}\log\sgn(\alpha-1)\sup_{p\in\M_f(\K)}\min_{\sigma\in\S(\hil)}g_{\alpha,0}(p,\sigma)\\
&\ds=
\frac{1}{\alpha-1}\log\sgn(\alpha-1)\min_{\sigma\in\S(\hil)}\sup_{\rho\in\K}\qrsr{\rho}{\sigma}{\alpha}\\
&\ds=
\min_{\sigma\in\S(\hil)}\sup_{\rho\in\K}\frac{1}{\alpha-1}\log\sgn(\alpha-1)\qrsr{\rho}{\sigma}{\alpha}\\
&\ds=
R_{S_\alpha}(\ran\ch).
\end{align*}
\end{IEEEproof}

\section{The one-shot classical capacity of quantum channels}\label{sec:one-shot capacities}

Let $W:\,\X\to\S(\hil)$ be a classical-quantum channel. In order to transmit (classical) information through the channel, the sender has to encode the messages into signals at the input of the channel, and the receiver has to make a measurement at the outcome to determine which message was sent. A \ki{code} is a triple $(M,\vfi,E)$, where $\{1,\ldots,M\}$ labels the possible messages to transmit, $\vfi:\,\{1,\ldots,M\}\to\X$ is the encoding map, and the positive operator valued measurement $E:\,\{1,\ldots,M\}\to\B(\hil)_+,\,\sum_{i=1}^M E_i=I$, is the decoding. 
The average probability of an erroneous decoding is given by 
\begin{equation*}
P_e(M,\vfi,E):=\frac{1}{M}\sum_{i=1}^M(1-\Tr W_{\vfi(i)}E_i)=1-P_s(M,\vfi,E),
\end{equation*}
where $P_s(M,\vfi,E)$ is the success probability. 
The one-shot $\ep$-capacity of the channel is defined as the logarithm of the maximal number of messages that can be transmitted through the channel with error not exceeding $\ep$:
\begin{align*}
C_\ep(\ch):=\max\{&\log M\,|\,\exists (M,\vfi,E)\s \text{such that}\\
& P_e(M,\vfi,E)\le\ep\}.
\end{align*}


Let $\chi^*_{H_r,0}(\ch)$ and $\chi^*_{S_\alpha,0}(\ch)$ denote the generalizations of the 
Holevo capacity of $\ch$ as 
defined in \eqref{def:cap0}, for a Hoeffding distance with parameter $r$ and for a R\'enyi 
relative entropy with parameter $\alpha$, respectively.
For any $\ep>0$ and any $c>0$, 
the one-shot $\ep$-capacity can be lower bounded as 
\begin{align*}
C_\ep(\ch)\ge& \chi^*_{H_{\log\bz(1+c)/\ep\jz},0}(W)-\log\bz\frac{2+c+1/c}{\ep}\jz\\
=&
\sup_{0\le\alpha<1}
\left\{\frac{-\alpha\log\bz\frac{1+c}{\ep}\jz}{1-\alpha}+\chi^*_{S_\alpha,0}(\ch)\right\}\\
&-\log\bz\frac{2+c+1/c}{\ep}\jz,
\end{align*}
where the inequality was shown in \cite{MD}, and the identity is obvious from the definition 
\eqref{hdist def} of the Hoeffding distances. While this bound might be rather loose for one single use of the channel, it is asymptotically optimal in the sense that it yields the Holevo capacity as a lower bound on the optimal asymptotic transmission rate of the channel \cite{MD}.

In order to give an upper bound on the capacity, one has to find an upper bound on the success probability for any code $(M,\vfi,E)$ in terms of $M$. Such a bound was given in \cite{ON}, that we briefly outline below.
 Note that the function $x\mapsto x^{\frac{1}{\alpha}}$ is operator monotonic increasing for $\alpha\in [1,+\infty)$ and thus
$\channel{\vfi(k)}=(\channel{\vfi(k)}^\alpha)^{\frac{1}{\alpha}}\le
 \bz\sum_{m=1}^M\channel{\vfi(m)}^\alpha\jz^{\frac{1}{\alpha}}$.
 Hence, the average success probability is upper bounded as
 \begin{align}
 P_s(M,\vfi,E)&\le
 \frac{1}{M}\sum_{k=1}^M \Tr E_k\bz\sum_{m=1}^M\channel{\vfi(m)}^\alpha\jz^{\frac{1}{\alpha}}\nonumber\\
&=
\frac{1}{M} \Tr \bz\sum_{m=1}^M\channel{\vfi(m)}^\alpha\jz^{\frac{1}{\alpha}}\nonumber\\
&=
M^\frac{1-\alpha}{\alpha} \Tr \bz\sum_{m=1}^M\frac{1}{M}\channel{\vfi(m)}^\alpha\jz^{\frac{1}{\alpha}}\nonumber\\
&\le
M^\frac{1-\alpha}{\alpha}\sup_{p\in\M_f(\X)}
2^{\frac{\alpha-1}{\alpha}\chi_\alpha(p)},\label{ON inequality}
 \end{align}
where
 \begin{equation*}
 \chi_{\alpha}(p):=\frac{\alpha}{\alpha-1}\log\Tr\omega(p),\ds\ds\ds \omega(p):=\bz\sum_{x\in\X}p(x)\channel{x}^\alpha\jz^{\frac{1}{\alpha}}.
 \end{equation*}
As it was pointed out in \cite{KW,Sibson}, for any $\sigma\in\S(\hil)$ and $p\in\M_f(\X)$ we have
\begin{align}
&S_\alpha\bz\Exp_p\hat\ch\,\big|\big|\,\hat p\otimes\sigma\jz\nonumber\\
&\ds=
S_\alpha\bz\Exp_p\hat\ch\,\Big|\Big|\,\hat p\otimes\frac{\omega(p)}{\Tr\omega(p)}\jz
+S_\alpha\bz\frac{\omega(p)}{\Tr\omega(p)}\,\Big|\Big|\,\sigma\jz\nonumber\\
&\ds=
\chi_\alpha(p)+S_\alpha\bz\frac{\omega(p)}{\Tr\omega(p)}\,\Big|\Big|\,\sigma\jz,
\label{formula:Renyi capacity}
\end{align}
and hence 
\begin{equation}\label{chi definition}
\chi_\alpha(p)=\inf_{\sigma\in\S(\hil)}\rsr{\Exp_p\hat\ch}{\hat p\otimes\sigma}{\alpha},
\end{equation}
which in turn yields 
\begin{equation}\label{capacity formula}
\sup_{p\in\M_f(\X)}\chi_\alpha(p)=\chi^*_{S_\alpha,1}(\ch).
\end{equation}

The above observations lead to the following:
 \begin{thm}\label{thm:one-shot upper bound}
 For any $\ep>0$, we have
  \begin{equation*}
 C_\ep(W)\le\inf_{\alpha>1}\left\{\chi_{S_\alpha,1}^*(W)+\frac{\alpha}{\alpha-1}\log\frac{1}{1-\ep}\right\}.
 \end{equation*}
 \end{thm}
 \begin{IEEEproof}
 Assume that for a code $(M,\vfi,E)$ we have $P_e(M,\vfi,E)\le \ep$. Then, by the above,
 \begin{equation*}
 \log(1-\ep)\le\log P_s(M,\vfi,E)
 \le\frac{\alpha-1}{\alpha}\bz \chi_{S_\alpha,1}^*(W)-\log M\jz
 \end{equation*}
for every $\alpha>1$,
 from which the assertion follows immediately. 
 \end{IEEEproof}


For each $n\in\N$, consider 
the $n$th i.i.d.~extension of $W$, defined as
$W^{(n)}:\,\X^n\to\S(\hil^{\otimes n})$,
\begin{equation*}
W^{(n)}(x_1,\ldots,x_n):=W(x_1)\otimes\ldots\otimes W(x_n).
\end{equation*}
The rate $R(\C)$ of a sequence of codes $\C=\{C^{(n)}=(M^{(n)},\vfi^{(n)},E^{(n)})\}_{n\in\N}$ is 
$R(\C):=\liminf_{n\to\infty}\frac{1}{n}\log M^{(n)}$, and 
the \ki{asymptotic $\ep$-capacity} of $W$ (with product encoding) is defined as
\begin{equation*}
\overline{C}_{\ep}(W):=\sup\big\{R(\C)\,\big|\,
\limsup_{n\to\infty} P_{e}(C^{(n)})\le\ep\big\},
\end{equation*}
where the supremum is taken over sequences of codes satisfying the indicated criterion. One can easily see that 
\begin{align*}
\liminf_{n\to\infty}\frac{1}{n}C_{\ep}(W^{(n)})&\le\overline{C}_{\ep}(W)\le \overline{C}_{\ep'}(W)\nonumber\\
&\le\liminf_{n\to\infty}\frac{1}{n}C_{\ep''}(W^{(n)})\label{ineq:capacity}
\end{align*}
for any $0\le \ep\le\ep'<\ep''$. 
The upper bound in Theorem \ref{thm:one-shot upper bound} is asymptotically sharp in the sense that it yields the Holevo capacity as an upper bound on the optimal information carrying capacity in the asymptotic limit. The details of the proof of the following Theorem are supplied in Appendix \ref{appendix:limit}.

\begin{thm}\label{thm:asymptotic capacity}
Assume that $\ran W$ is compact. Then,
for any $\ep\in[0,1)$,
\begin{equation*}
\overline{C}_{\ep}(\ch)\le\chi_S^*(\ch).
\end{equation*}
\end{thm}
\begin{IEEEproof}
By Theorem \ref{thm:one-shot upper bound} and Proposition \ref{prop:capacity additivity},
\begin{align*}
\overline{C}_{\ep}(W)&\le\liminf_{n\to\infty}\frac{1}{n}C_{\ep'}(W^{(n)})\\
&\le\liminf_{n\to\infty}\left\{\frac{1}{n}\chi_{S_\alpha,1}^*(W^{(n)})
+\frac{1}{n}\frac{\alpha}{\alpha-1}\log\frac{1}{1-\ep'}\right\}\\
&=\chi_{S_\alpha,1}^*(W)
\end{align*}
for any $0<\ep<\ep'<1$ and $\alpha>1$. 
By Proposition \ref{prop:capacity limit}, the assertion follows for every $\ep>0$, and
the case $\ep=0$ is immediate from $\overline{C}_{0}(\ch)\le\overline{C}_{\ep}(\ch),\,\ep>0$.
\end{IEEEproof}
\medskip

\begin{rem}\label{rem:channel cutoff rates}
Cutoff rates were also defined in \cite{Csiszar} for channel coding in the following way:
for $\kappa<0$, the $\kappa$-cutoff rate $C_\kappa(W)$ is the largest $R_0$ for which
\begin{equation*}
\limsup_{n\to\infty}\frac{1}{n}\log P_e(C^{(n)})\le \kappa(R_0-R)
\end{equation*}
for any sequence of codes with rate $R$, while for $\kappa>0$, the $\kappa$-cutoff rate $C_\kappa(W)$ is the largest $R_0$ for which
\begin{equation*}
\limsup_{n\to\infty}\frac{1}{n}\log P_s(C^{(n)})\le \kappa(R_0-R)
\end{equation*} 
for any sequence of codes with rate $R$.

Inequality \eqref{ON inequality} and identity \eqref{capacity formula}, together with the observations of Appendix \ref{appendix:limit}, yield that, for $\alpha>1$,
\begin{equation*}
\limsup_{n\to\infty}\frac{1}{n}\log P_s(C^{(n)})\le\frac{\alpha-1}{\alpha}(\chi^*_{S_\alpha,1}(W)-R)
\end{equation*}
for any sequence of codes with rate $R$ and hence,
\begin{equation*}
C_\kappa(W)\ge \chi^*_{S_{\frac{1}{1-\kappa}},1}(W),\ds\ds\ds 0<\kappa<1.
\end{equation*}
The above inequality was shown to hold as an equality for classical channels in \cite{Csiszar}.
\end{rem}

\section{Remarks on the divergence radius}

Let $\Sigma$ be a subset of the state space $\S(\hil)$, and let $R_D(\Sigma)$ denote its $D$-radius as given in \eqref{divergence radius}.
A state $\sigma^*$ which reaches the infimum in \eqref{divergence radius} is called a \ki{$D$-centre} for $\Sigma$. As we have seen in the previous section, the $S_\alpha$-radii of the range of a channel are related to the direct part of channel coding for $\alpha\in[0,1)$ and to the converse part for $\alpha\in(1,+\infty]$. In both cases, the asymptotically relevant quantities are the divergence radii with $\alpha$ close to $1$. On the other hand, for state discrimination the relevant quantity turns out to be the $\infty$-radius. More precisely, if $\rho_1,\ldots,\rho_r\in\S(\hil)$ then the optimal success probability of discriminating them by POVM measurements is given by $P_s=(1/r)\exp\bz R_{S_{\max}}\{\rho_k\}\jz$ \cite{KRS}, where $S_{\max}$ is the max-relative entropy \cite{Datta}. 

Related to state discrimination is the following geometrical problem: given $\rho_1,\ldots,\rho_r\in\S(\hil)$, find the largest $q$ such that there exist states $\tau_1,\ldots,\tau_r$ such that 
$q\rho_i+(1-q)\tau_i$ is independent of $i$. Such a family of states $\tau_1,\ldots,\tau_r$ is called an optimal Helstr\"om family with parameter $q$ in \cite{KMI}. As one can easily see, the largest such $q$ is given by $\exp\bz-R_{S_{\max}}\{\rho_k\}\jz$, and $q\rho_i+(1-q)\tau_i$ is an $S_{\max}$-centre for $\{\rho_k\}_{k=1}^r$.
When $r=2$,
the results of Holevo \cite{Holevo} and Helstr\"om \cite{Helstrom} yield that the optimal success probability is given by $P_s=(1+D)/2$, where $D:=(1/2)\norm{\rho_1-\rho_2}_1$, and hence,
$R_{S_{\max}}(\{\rho_1,\rho_2\})=\log(1+D)$.
Moreover, an $S_{\max}$-centre is given by 
$\sigma^*=(\rho_1+2X_+)/(1+D)=(\rho_2+2X_-)/(1+D)$, where $X_+$ and $X_-$ are the positive and the negative parts of $\rho_1-\rho_2$, respectively. 
In \cite{AF} and \cite{HSR},
a suboptimal Helstr\"om family was used for two states $\rho_1$ and $\rho_2$ to show Fannes type inequalities. Using instead the above optimal Helstr\"om family in the proof of \cite[Proposition 1]{HSR}, one obtains the following:
\begin{prop}
Let $\hil$ be a Hilbert space and $f:\,\S(\hil)\to \iC$ be a bounded function that satisfies
\begin{equation}\label{mixture condition}
\abs{f((1-\ep)\rho_1+\ep\rho_2)-(1-\ep)f(\rho_1)-\ep f(\rho_2)}\le h_2(\ep)
\end{equation}
for any two states $\rho_1,\rho_2$ and any $\ep\in[0,1]$, where $h_2(x):=-x\log x-(1-x)\log(1-x)$ is the binary entropy function.
Then, for any two states $\rho_1,\rho_2$ on $\hil$, we have
\begin{equation}\label{Fannes inequality}
|f(\rho_1)-f(\rho_2)|\le 2h_2(\ep)+4\ep M,
\end{equation}
where $\ep:=\frac{\norm{\rho_1-\rho_2}_1}{2+\norm{\rho_1-\rho_2}_1}$ and $M:=\sup_{\rho\in\S(\hil)}|f(\rho)|$. 
\end{prop}
\begin{IEEEproof}
Let $\tau_1,\tau_2$ be the above optimal Helstr\"om family and $\sigma^*=(1-\ep)\rho_i+\ep\tau_i$ be the $S_{\max}$-centre of $\{\rho_1,\rho_2\}$. Then,
\begin{align*}
&|f(\rho_1)-f(\rho_2)|\\
&\ds\le |f(\rho_1)-f(\sigma^*)|+|f(\sigma^*)-f(\rho_2)|\\
&\ds\le
\sum_{i=1}^2|f(\sigma^*)-(1-\ep)f(\rho_i)-\ep f(\tau_i)|+\ep|f(\rho_i)|+\ep|f(\tau_i)|\\
&\ds\le
2h_2(\ep)+4\ep M.
\end{align*}
\end{IEEEproof}

The von Neumann entropy is known to satisfy \eqref{mixture condition}, which in turn yields by a simple computation that 
the conditional entropy and the relative entropy distance from a convex set containing a faithful state satisfy \eqref{mixture condition}, too. Note that for the latter two quantities \eqref{Fannes inequality} yields a slight improvement of the result of \cite{AF} and of \cite[Lemma 1]{HSR}, respectively, where the same bound was obtained with $\ep=\norm{\rho_1-\rho_2}_1$.

For the case where $D$ is the relative entropy $S$, it was shown in \cite{OPW} that for any subset $\Sigma$ of states, the $S$-centre is unique and is inside the closed convex hull $\overline{\mathrm{co}}\Sigma$ of $\Sigma$. This is no longer true for other R\'enyi relative entropies in general. For instance, for the classical probability distributions $\rho_1:=(1/2,1/4,1/4),\, \rho_2:=(1/2,1/6,1/3)$, an $S_\infty$-centre is given by $\sigma^*=(6/13, 3/13, 4/13)$, and one can easily verify that no $S_\infty$-centre can be found on the line segment connecting $\rho_1$ and $\rho_2$. It is of some mathematical interest to find conditions on $D$ ensuring the existence of a unique $D$-centre of $\Sigma$ in $\overline{\mathrm{co}}\Sigma$ for any subset of states $\Sigma$.


\section{Concluding remarks}

The idea of representing the R\'enyi relative entropies as cutoff rates is from Csisz\'ar 
\cite{Csiszar}, and we essentially followed his approach here. 
Note, however, that the analysis of the error exponents $\underline{h}_r,\overline{h}_r,h_r$ 
in the classical case, on which the proof of \cite{Csiszar} relies, is based on the 
Hellinger arc and a representation of the Hoeffding distances that have no equivalents in 
the quantum setting \cite{ON2}. Instead, our analysis is based on an equivalent definition 
of the Hoeffding distances that can be defined also for quantum states, given in 
\eqref{hdist def}.
That this definition of the Hoeffding distances have the right operational meaning was 
proven recently under the name of the quantum Hoeffding bound 
\cite{ANSzV,Hayashi,HMO2,Nagaoka}. 
Note that this representation of the Hoeffding distances allows for a somewhat simplified 
proof even in the classical case. Moreover, this proof works also for the more general 
setting of correlated states considered in Theorem \ref{thm:HMO2}.

The way to prove the identity of the different definitions of the R\'enyi capacities using minimax results is also from \cite{Csiszar}. For this, the convexity of $\sigma\mapsto\qrsr{\rho}{\sigma}{\alpha}$ and $\sigma\mapsto\rsr{\rho}{\sigma}{\alpha}$ for every fixed $\rho$ are essential. These are obvious in the classical case for $Q_\alpha$, and for $S_\alpha$ when
$\alpha\in (0,1)$, and were proven for $S_\alpha$ and $\alpha>1$ in \cite{Csiszar}. 
That proof, however, cannot be extended to the quantum case and, as far as we are aware, our Theorem \ref{thm:convexity} is a new result. Note that in the quantum case the fact that $x\mapsto x^{1-\alpha}$ is not operator convex for $\alpha>2$ yields a strong limitation, and no convexity properties of the $\alpha$-relative entropies are expected to hold for parameters $\alpha>2$. 
This limitation was overcome in \cite{KW}, where a completely different approach was used to prove that $\chi_{S_\alpha,1}^*=R_{S_\alpha}(\ran W)$ for all $\alpha>1$.
Another subtle technical difference between the proofs for the classical (more precisely, finite $\X$) and the general 
cases comes from the fact that in minimax theorems one of the sets has to be compact and 
convex, which in the first case can be chosen to be $\M_f(\X)$, 
and the other space has to be convex, which is chosen to be $\S(\hil)_{++}$.
In the general case $\X$ is usually the state space of a quantum system, which is of  
infinite cardinality and hence $\M_f(\X)$ is convex but not compact, whereas replacing 
$\M_f(\X)$ with $\M_m(\ran W)$ as in Appendix \ref{appendix:limit} yields a space that is 
compact but not convex. Hence we switched the role of the two spaces and chose $\S(\hil)$ to 
be the compact convex set. However, the (dis)continuity properties of the R\'enyi relative 
entropies then wouldn't make it possible to satisfy the continuity requirements of minimax theorems, and that's why we had to use $\ep$-perturbations in Section \ref{sec:equivalence}.

It is worth noting that R\'enyi relative entropies and the corresponding channel capacities 
are related to different regimes of information-theoretic tasks for the parameter values 
$\alpha\in(0,1)$ and for $\alpha\in(1,+\infty)$. Indeed, the first interval is related to 
the so-called direct part of problems, i.e., where a relevant error probability decays 
exponentially for rates below the optimal one, while the second interval is related to the 
(strong) converse regions, where a relevant success probability goes to zero (exponentially) 
for rates above the optimal rate. 
Cutoff rates are also defined in an asymmetric way, separately for the direct region ($\kappa<0$) and for the strong converse region $(\kappa>0)$;
see Remark \ref{rem:channel cutoff rates} 
and \cite{Csiszar} for more details.

In the case of hypothesis testing between $\rho$ and $\sigma$, 
for rates $r<\sr{\sigma}{\rho}$, 
the optimal exponential decay rates of the error probabilities of the second kind are given 
explicitly by the Hoeffding distances $\hdist{\rho}{\sigma}{r}$, which are defined through the 
R\'enyi relative entropies $\rsr{\rho}{\sigma}{\alpha},\,\alpha\in(0,1)$.
For rates 
$r>\sr{\sigma}{\rho}$, the success probabilities decay exponentially, and the optimal decay 
rates are known in the classical case to be given by the Han-Kobayashi bounds 
\cite{HK,ON2,Hayashi_book}, defined through 
$\rsr{\rho}{\sigma}{\alpha},\,\alpha\in(1,+\infty)$.
In the quantum case, however, the exact error exponents for the converse 
part are not known and hence it is not possible to extend the results of \cite{Csiszar} on the
cutoff rates for $\kappa>0$ at the moment, though the 
results of \cite{ON2,Hayashi_book} give inequalities between the cutoff rates and the R\'enyi relative 
entropies that are expected to hold as equalities.
For channel coding, the exact error exponents are not known for every rate value even in 
the classical case, but we see the same picture, i.e., the exponential decay of error 
probabilities for rates below the Shannon capacity can be expressed in terms of, or upper 
bounded by, the R\'enyi capacities $\chi^*_{S_\alpha}$ with $\alpha\in(0,1)$, while 
for rates above the Shannon capacity, the 
exponential decay rate of success probabilities can be expressed 
in terms of the R\'enyi capacities $\chi^*_{S_\alpha}$ with $\alpha\in(1,+\infty)$ \cite{Csiszar}.

Due to finite-size effects, the one-shot capacities are discontinuous functions of the error bar $\ep$, and they depend on the parameters of the channel in a more intricate way than their asymptotic counterparts. As a result, it doesn't seem to be likely that they could be expressed in a similarly compact form as the asymptotic capacities, and 
if one is looking for some universal statement on them, applicable to all channels and all possible error bars, then probably the best one can hope for are lower and upper estimates on their values.
In view of the above noted difference between the role of the intervals $\alpha\in(0,1)$ and $\alpha\in(1,+\infty)$, it seems rather natural to expect lower bounds in terms of the capacities $\chi^*_{S_\alpha}$ with $\alpha\in(0,1)$ and upper bounds in terms of 
the capacities $\chi^*_{S_\alpha}$ with $\alpha\in(1,+\infty)$. While 
we left the question of optimality open for the bounds provided in 
Section \ref{sec:one-shot capacities} (in fact, even to formulate what optimality might mean in this setting is a non-trivial question), it is somewhat reassuring that the optimal asymptotic capacity can be recovered by applying our bounds to several copies of the channel and 
letting the number of copies go to infinity.

%

\appendices
\section{A minimax theorem}\label{appendix:minimax}

Let $X$ and $Y$ be non-empty sets and $f:\,X\times Y\to \overline{\R}:=\R\cup\{-\infty,+\infty\}$ be a function.
Obviously, for any $x_0\in X$ and $y_0\in Y$ we have 
$\inf_{x\in X} f(x,y_0)\le f(x_0,y_0)\le\sup_{y\in Y} f(x_0,y)$ and hence,
\begin{equation}\label{minimax ineq}
\sup_{y\in Y}\inf_{x\in X}f(x,y)\le\inf_{x\in X}\sup_{y\in Y}f(x,y).
\end{equation} 
Minimax theorems give sufficient conditions on when the above inequality holds with equality. 
The following Lemma \ref{lemma:minimax finiten} is a step in the proof of Sion's minimax theorem in \cite{Komiya}, the proof of which we include for the readers' convenience.
We will use the notation $[f(.\,,y)\le c]$ to denote the level set $\{x\in X\,:\,f(x,y)\le c\}$ for some number $c\in \R$, and other level sets are denoted similarly. 

\begin{lemma}\label{lemma:minimax finiten}
Assume that $X$ is a compact topological space and $f(.\,,\,y)$ is lower semicontinuous for every $y\in Y$. Assume, moreover, that for any finite subset $Y'\subset Y$ we have
\begin{equation}\label{minimax finiten}
\inf_{x\in X}\max_{y\in Y'} f(x,y)\le \sup_{y\in Y}\inf_{x\in X}f(x,y).
\end{equation}
Then the infima in \eqref{minimax ineq} can be replaced with minima, and
\begin{equation*}
\sup_{y\in Y}\min_{x\in X}f(x,y)=\min_{x\in X}\sup_{y\in Y}f(x,y).
\end{equation*} 
\end{lemma}
\begin{IEEEproof}
The lower semi-continuity of $f(.\,,y),\,y\in Y$ implies the 
lower semi-continuity of $\sup_y f(.\,,y)$ and, since $X$ is compact,
all the functions $f(.\,,y),\,y\in Y$, and $\sup_y f(.\,,y)$ reach their infima on $X$. Hence, we can replace the infima with minima. 

To prove the main assertion, we have to show that 
\begin{equation*}
\min_{x\in X}\sup_{y\in Y}f(x,y)\le\sup_{y\in Y}\min_{x\in X}f(x,y).
\end{equation*} 
Let $c<\min_{x\in X}\sup_{y\in Y}f(x,y)$ or equivalently, let $c$ be such that 
$\bigcap_{y\in Y}[f(.\,,y)\le c]=\emptyset$.
Lower semicontinuity of $f(.\,,y)$ yields that $[f(.\,,y)\le c]$ is closed (and hence compact) for every $y\in Y$ and hence, there exist finitely many $y_1,\ldots,y_r$ such that 
$\bigcap_{i=1}^r[f(.\,,y_i)\le c]=\emptyset$ or equivalently, 
$c<\min_{x\in X}\max_{1\le i\le r} f(x,y_i)$. By the assumption \eqref{minimax finiten}, we obtain
$c<\sup_{y\in Y}\min_{x\in X}f(x,y)$. Since this holds for any $c<\min_{x\in X}\sup_{y\in Y}f(x,y)$, the assertion follows.
\end{IEEEproof}

\begin{cor}\label{cor:minimax}
Let $X$ be a compact topological space, $Y$ be a subset of the real line and let $f:\,X\times Y\to \overline{\R}$ be a function. Assume that 
\begin{enumerate}
\item
$f(.\,,\,y)$ is lower semicontinuous for every $y\in Y$ and
\item
$f(x,.)$ is monotonic increasing for every $x\in X$, or
$f(x,.)$ is monotonic decreasing for every $x\in X$.
\end{enumerate}
Then the infima in \eqref{minimax ineq} can be replaced with minima, and
\begin{equation*}
\sup_{y\in Y}\min_{x\in X}f(x,y)=\min_{x\in X}\sup_{y\in Y}f(x,y).
\end{equation*} 
\end{cor}
\begin{IEEEproof}
By the monotonicity assumption,
for any finite subset $Y'=\{y_1,\ldots,y_r\}\subset Y$, there exists a $y^*\in\{y_1,\ldots,y_r\}$ such that 
\begin{equation*}
\max_{1\le i\le r} f(x,y_i)=f(x,y^*)
\end{equation*}
for all $x\in X$. Hence,
\begin{equation*}
\min_{x\in X}\max_{1\le i\le r} f(x,y_i)=
\min_{x\in X}f(x,y^*)
\le \sup_{y\in Y}\min_{x\in X}f(x,y).
\end{equation*}
Thus, all the conditions of Lemma \ref{lemma:minimax finiten} are satisfied, from which the assertion follows.
\end{IEEEproof}

\section{The limit of the $\alpha$-capacities}
\label{appendix:limit}

In this Appendix we collect some properties of the quantities $\chi_\alpha$ and $\chi_\alpha^*$ that are needed for the proof of Theorem 
\ref{thm:asymptotic capacity}. To simplify notation, we introduce
\begin{equation*}
\chi^*_{S_\alpha}:=\chi^*_{S_\alpha,1}=\chi^*_{S_\alpha,2}=R_{S_\alpha}(\ran W),
\end{equation*}
where $W:\,\X\to\S(\hil)$ is a fixed classical-quantum channel.

We start with the following:
\begin{lemma}\label{lemma:concavity}
Assume that $\alpha>1$. Then, for any $p_1,p_2\in\M_f(\X)$, $\eta\in(0,1)$ and $\sigma\in\S(\hil)$,
 \begin{align}
&\rsr{\Exp_{(1-\eta)p_1+\eta p_2}\hat\ch}{((1-\eta)\hat p_1+\eta \hat p_2) \otimes\sigma}{\alpha}\label{chi concavity1}\\
&\ge
(1-\eta)\rsr{\Exp_{p_1}\hat\ch}{\hat p_1 \otimes\sigma}{\alpha}+
\eta\rsr{\Exp_{p_2}\hat\ch}{\hat p_2 \otimes\sigma}{\alpha}\label{chi concavity2}\\
&\ge
(1-\eta)\chi_{_\alpha}(p_1)+\eta\chi_{_\alpha}(p_2).\label{chi concavity3}
\end{align}
In particular,
the function $p\mapsto\chi_\alpha(p)$ is concave on $\M_f(\X)$.
\end{lemma}
\begin{IEEEproof}
The inequality in \eqref{chi concavity3} is obvious from \eqref{formula:Renyi capacity}.
One can easily verify that the expression in \eqref{chi concavity1} is equal to $+\infty$ if and only if the expression in \eqref{chi concavity2} is equal to $+\infty$, and otherwise the inequality between the two follows by a straightforward computation from the 
concavity of the function $\frac{1}{\alpha-1}\log$. The last assertion follows by taking the infimum in $\sigma$ in the inequality between \eqref{chi concavity1} and \eqref{chi concavity3}.
\end{IEEEproof}
\medskip

The following statement is essentially Lemma 2 from \cite{ON}:
\begin{prop}\label{prop:capacity additivity}
Assume that $\ran W$ is compact and $\alpha>1$. Then
\begin{equation*}
\chi^*_{S_\alpha}(W^{(n)})=n\chi^*_{S_\alpha}(W),\ds\ds\ds n\in\N.
\end{equation*}
\end{prop}
\begin{IEEEproof}
 Using the concavity established in Lemma \ref{lemma:concavity}, one can 
 follow the proof of Lemma 2 in \cite{ON} to obtain the assertion.
 (Note that in \cite{ON}, $\X$ was assumed to be finite, but that doesn't make a difference in the proof.) 
\end{IEEEproof}
\medskip

Let $m:=(\dim\hil)^2+1$, and let
\begin{equation*}
\M_m(\ran\ch):=\{p\in\M_f(\ran\ch)\,:\,|\supp p|\le m\}
\end{equation*}
denote the set of probability measures supported on not more than $m$ points in $\ran\ch$.
By Carath\'eodory's theorem \cite[Theorem (2.3)]{Barvinok}, for every $p\in\M_f(\X)$, there exists a $\tilde p\in \M_{m}(\ran\ch)$ such that 
\begin{equation*}
\capt{\alpha}(p)=\tilde\chi_{\alpha}(\tilde p):=\frac{\alpha}{\alpha-1}\log\Tr\bigg(\sum_{\omega\in\ran \ch}\tilde p(\omega)\omega^{\alpha}\bigg)^{\frac{1}{\alpha}}.
\end{equation*}
Note that $\tilde\chi$ can also be defined by replacing $\X$ with $\ran W$ and $W$ with the identity map $\id$ on $\ran W$ in \eqref{chi definition}, i.e., for each $p\in\M_f(\ran W)$,
\begin{equation}\label{tilde chi definition}
\tilde\chi_{\alpha}(p)=\inf_{\sigma\in\S(\hil)}\rsr{\Exp_p\hat\id}{\hat p\otimes\sigma}{\alpha}.
\end{equation}
The functions $\chi_1$ and $\tilde\chi_1$ are defined simply by replacing $\alpha$ with $1$ in \eqref{chi definition} and in \eqref{tilde chi definition}, respectively.
Note that 
\begin{equation*}
\chi_{S_\alpha}^*(W)=\sup_{p\in\M_f(\X)}\chi_\alpha(p)=\sup_{p\in\M_m(\ran W)}\tilde\chi_\alpha(p)
\end{equation*}
for every $\alpha\in [0,+\infty)$.

\begin{lemma}\label{lemma:capacity limit}
The functions $\alpha\mapsto \chi_\alpha(p)$ and $\alpha\mapsto \tilde\chi_\alpha(p)$ are montonically increasing on $[0,+\infty)$ for all $p\in\M_f(\X)$ and $p\in\M_f(\ran W)$, respectively, and
\begin{align*}
\lim_{\alpha\to 1}\chi_\alpha(p)=\chi_{1}(p),\ds\ds\ds
\lim_{\alpha\to 1}\tilde\chi_\alpha(p)=\tilde\chi_{1}(p).
\end{align*}
\end{lemma}
\begin{IEEEproof}
The assertion on the monotonicity follows immediately from the monotoncity of the R\'enyi relative entropies in the parameter $\alpha$. We prove the assertion on the limit 
separately for $\alpha\nearrow 1$ and for $\alpha\searrow 1$. 
In the second case, we have
\begin{align*}
\lim_{\alpha\searrow 1}\chi_\alpha(p)&=
\inf_{\alpha>1}\chi_\alpha(p)=
\inf_{\alpha>1}\inf_{\sigma\in\S(\hil)}\rsr{\Exp_p\hat\ch}{\hat p\otimes\sigma}{\alpha}\\
&=
\inf_{\sigma\in\S(\hil)}\inf_{\alpha>1}\rsr{\Exp_p\hat\ch}{\hat p\otimes\sigma}{\alpha}\\
&=
\inf_{\sigma\in\S(\hil)}\sr{\Exp_p\hat\ch}{\hat p\otimes\sigma}
=
\chi_{1}(p).
\end{align*}

For fixed $p\in\M_f(\X)$ and $\alpha\in [0,+\infty)$, the map 
$\sigma\mapsto\rsr{\Exp_p\hat\ch}{\hat p\otimes\sigma+\ep I}{\alpha}$
is continuous on the compact set $\S(\hil)$ and hence the map
$\sigma\mapsto\rsr{\Exp_p\hat\ch}{\hat p\otimes\sigma}{\alpha}$ is lower semicontinuous, due to 
\eqref{ep limit S}.
On the other hand, for fixed $\sigma\in\S(\hil)$, the map
$\alpha\mapsto \rsr{\Exp_p\hat\ch}{\hat p\otimes\sigma}{\alpha}$ is monotonic increasing in $\alpha$ and hence, by Corollary \ref{cor:minimax}, we have
\begin{align*}
\lim_{\alpha\nearrow 1}\chi_{_\alpha}(p)&=
\sup_{\alpha<1}\inf_{\sigma\in\S(\hil)}\rsr{\Exp_p\hat\ch}{\hat p\otimes\sigma}{\alpha}\\
&=
\inf_{\sigma\in\S(\hil)}\sup_{\alpha<1}\rsr{\Exp_p\hat\ch}{\hat p\otimes\sigma}{\alpha}\\
&=
\inf_{\sigma\in\S(\hil)}\sr{\Exp_p\hat\ch}{\hat p\otimes\sigma}
=
\chi_{1}(p).
\end{align*}
The proof for $\lim_{\alpha\to 1}\tilde\chi_\alpha(p)$ goes exactly the same way.
\end{IEEEproof}
\medskip

The following Lemma was shown in \cite{FN}. For readers' conveniance, we include a proof here.
\begin{lemma}\label{lemma:compact}
If $\ran W$ is compact then $\M_m(\ran W)$ can be equipped with a topology $\tau$ with respect to which 
$\M_m(\ran W)$ is compact and $\tilde\chi_\alpha$ is continuous.
\end{lemma}
\begin{IEEEproof}
Let 
$S_m:=\{(\lambda_1,\ldots,\lambda_m)\,:\,\lambda_1,\ldots,\lambda_m\ge 0,\,\sum_{i=1}^m\lambda_i=1\}$ denote the $m$-dimensional probability simplex, and define
$\Omega_m(\ch):=S_{m}\times(\ran W)^{m}=\left\{(\underline{\lambda},\underline{\omega})\,:\,
\underline{\lambda}\in S_{m},\,\omega_1,\ldots,\omega_{m}\in\ran\ch
\right\}$.
Compactness of $\ran \ch$ yields that $\Omega_m(\ch)$ is compact with respect to its natural topology. Let $\pi_m:\,\Omega_m(\ch)\to\M_m(\ran \ch)$, $\pi_m(\underline{\lambda},\underline{\omega}):=\sum_{i=1}^m\lambda_i\delta_{\omega_i}$, where $\delta_{\omega_i}$ denotes the Dirac measure concentrated at $\omega_i$.
We define the topology $\tau$ on $\M_m(\ran\ch)$ to be the factor topology, i.e., the finest topology with respect to which $\pi_m$ is continuous. Being the continuous image of a compact set, $\M_m(\ran\ch)$ is also compact. 
One can easily see that $\tilde\chi_\alpha\circ\pi_m$ is continuous on $\Omega_m(\ch)$, which in turn is equivalent to the continuity of $\tilde\chi_\alpha$ with respect to $\tau$.
\end{IEEEproof}
\medskip

The following statement was shown in Lemma 3 of \cite{ON} for the case where $\X$ is finite. Here we give an alternative proof, using the minimax theorem established in Appendix \ref{appendix:minimax}, that covers the general case.
\begin{prop}\label{prop:capacity limit}
\begin{equation*}
\lim_{\alpha\to 1}\chi^*_{S_\alpha}(W)=\chi_{S}^*(W).
\end{equation*}
\end{prop}
\begin{IEEEproof}
We prove separately the cases $\alpha\nearrow 1$ and $\alpha\searrow 1$. 
In the first case, the assertion follows immediately from Lemma \ref{lemma:capacity limit}, as
\begin{align*}
\lim_{\alpha\nearrow 1}\chi_{S_\alpha}^*(W)&=
\sup_{\alpha\in[0,1)}\sup_{p\in\M_f(\X)}\chi_{\alpha}(p)\\
&=
\sup_{p\in\M_f(\X)}\sup_{\alpha\in[0,1)}\chi_{\alpha}(p)\\
&=
\sup_{p\in\M_f(\X)}\chi_{1}(p)
=
\chi^*_{S}(W).
\end{align*}

Note that the function $f(p,\alpha):=-\tilde\chi_\alpha(p)$ is monotonic decreasing in its second variable on $Y:=(1,+\infty)$ and continuous in its first variable on the compact space $X:=\M_m(\ran W)$, due to Lemma \ref{lemma:compact}. Hence, we can apply the minimiax theorem of Corollary \ref{cor:minimax} to obtain
\begin{align*}
\lim_{\alpha\searrow 1}\chi_{S_\alpha}^*(W)&=
\inf_{\alpha>1}\max_{p\in\M_m(\ran W)}\tilde\chi_{\alpha}(p)\\
&=
\max_{p\in\M_m(\ran W)}\inf_{\alpha>1}\tilde\chi_{\alpha}(p)\\
&=
\max_{p\in\M_m(\ran W)}\tilde\chi_{1}(p)
=
\chi^*_{S}(W).
\end{align*}

\end{IEEEproof}

\section*{Acknowledgments}
M. Mosonyi is grateful to Tomohiro Ogawa and Andreas Winter for helpful discussions.

\ifCLASSOPTIONcaptionsoff
  \newpage
\fi

\end{document}